\documentclass[aps,prc,twocolumn,longbibliography,reprint,superscriptaddress,nofootinbib]{revtex4-1}
\usepackage[colorlinks=true,urlcolor=blue,linkcolor=blue,anchorcolor=blue,citecolor=red]{hyperref}
\usepackage[utf8]{inputenc}
\usepackage{natbib}
\usepackage{graphicx,color}
\usepackage{amsmath,amsfonts,amsthm,amssymb,mathrsfs}

\graphicspath{{fig/}}

\begin{document}

\preprint{}

\title{Hydrodynamical attractor and thermal particle production in heavy-ion collision}

\author{Lakshmi J. Naik}
\email{jn\_lakshmi@cb.students.amrita.edu}
\affiliation{Department of Sciences, Amrita School of Engineering, Coimbatore, Amrita Vishwa Vidyapeetham, India}
\author{Sunil Jaiswal}
\email{sunil.jaiswal@tifr.res.in}
\affiliation{Department of Nuclear and Atomic Physics, Tata Institute of Fundamental Research, Mumbai 400005, India}
\author{K. Sreelakshmi}
\affiliation{Department of Sciences, Amrita School of Engineering, Coimbatore, Amrita Vishwa Vidyapeetham, India}
\author{Amaresh Jaiswal}
\email{a.jaiswal@niser.ac.in}
\affiliation{School of Physical Sciences, National Institute of Science Education and Research, HBNI, Jatni 752050, Odisha, India}
\author{V. Sreekanth}
\email{v\_sreekanth@cb.amrita.edu}
\affiliation{Department of Sciences, Amrita School of Engineering, Coimbatore, Amrita Vishwa Vidyapeetham, India}
\date{\today}
%


\begin{abstract}
Study of thermal particle production is crucial to understand the space-time evolution of the fireball produced in high energy heavy-ion collisions. We consider thermal particle production within the framework of relativistic viscous hydrodynamics and employ recently obtained analytical solutions of higher-order viscous hydrodynamics with longitudinal Bjorken expansion to calculate the spectra of dileptons and photons. Using these analytical solutions, we constrain the allowed initial states by demanding positivity and reality of energy density throughout the evolution. Further, we compute thermal particle spectra and study the particle yield in context of hydrodynamic attractors. We find that, of all allowed solutions, the evolution corresponding to attractor solution leads to maximum production of thermal particles. 
\end{abstract}



\maketitle


\section{Introduction}
\label{intro}

High energy nucleus-nucleus collisions at the Relativistic Heavy Ion Collider (RHIC) and the Large Hadron Collider (LHC) produce rapidly expanding, strongly interacting QCD matter, which consists of deconfined states of quarks and gluons over nuclear volume known as the quark-gluon plasma (QGP). Relativistic hydrodynamic simulations have been extremely successful in describing the space-time evolution of QGP formed in the early stages of collisions; see Refs.~\cite{Shuryak:2003xe, Ollitrault:2012cm, Heinz:2013th, Gale:2013da, Jaiswal:2016hex} for detailed review. The causal description of this expanding viscous medium has created a lot of renewed interest in different arena including the formal development of relativistic dissipative hydrodynamics framework; see Refs.~\cite{Florkowski:2017olj, Romatschke:2017ejr} for review of recent theoretical activities.

The formulation of relativistic dissipative hydrodynamics usually proceeds with the assumption of the system being close to local thermodynamic equilibrium. On the other hand, it was demonstrated that hydrodynamic simulations were quite successful in explaining flow data in small collision systems (which are expected to be far from equilibrium) formed in proton-proton and proton-lead collisions \cite{Bozek:2015swa, Bozek:2016jhf}. This unexpected success of hydrodynamics has recently generated much interest in the foundational aspects of causal theory of relativistic fluid dynamics \cite{Heller:2013fn, Heller:2015dha, Aniceto:2015mto, Basar:2015ava, Florkowski:2016zsi, Heller:2016rtz, Romatschke:2016hle, Behtash:2017wqg, Blaizot:2017lht, Blaizot:2017ucy, Denicol:2017lxn, Romatschke:2017acs, Romatschke:2017vte, Strickland:2017kux, Spalinski:2017mel, Aniceto:2018uik, Behtash:2018moe, Chattopadhyay:2018apf, Denicol:2018pak, Heller:2018qvh, Strickland:2018ayk, Tinti:2018qfb, Behtash:2019txb, Blaizot:2019scw, Grozdanov:2019uhi, Grozdanov:2019kge, Heinz:2019dbd, Strickland:2019hff, Jaiswal:2019cju, Denicol:2019lio, Blaizot:2020gql, Jaiswal:2020hvk}. In the current work, we consider an intriguing feature that manifests in boost-invariant causal theories of relativistic viscous fluid dynamics, ``the hydrodynamic attractor" \cite{Heller:2015dha, Romatschke:2017acs, Denicol:2018pak, Behtash:2017wqg, Strickland:2017kux, Chattopadhyay:2018apf}, and study its phenomenological implications via thermal particle production.

Attractor feature is observed when one considers the hydrodynamic formulation as an expansion in gradient series. However, the gradient expansion generates an asymptotic series with zero radius of convergence \cite{Heller:2015dha, Basar:2015ava}. Fortunately, this diverging series can be Borel resummed resulting in a distinctive hydrodynamic attractor solution (the hydrodynamic mode), together with an infinite set of rapidly decaying non-hydrodynamic modes that specifies the approach towards this attractor solution starting from arbitrary initial conditions \cite{Heller:2013fn, Heller:2015dha}. This indicates that hydrodynamic solutions exhibit a certain degree of universality resulting from rapid memory loss of initial conditions. This feature of hydrodynamic attractor has been used in recent works to establish a macroscopic description of the out-of-equilibrium dynamics of high energy heavy-ion collisions \cite{Giacalone:2019ldn, Coquet:2021cuv}. Recently, for a system with one dimensional longitudinal boost-invariant expansion \cite{Bjorken:1982qr}, the solutions, including the attractor solution, of causal hydrodynamic evolution equations were analytically obtained for several physically relevant cases \cite{Denicol:2017lxn, Jaiswal:2019cju}. It is therefore of interest to consider these analytical solutions as well as attractors to study observables relevant to relativistic heavy-ion collisions.

Thermal particles, such as dileptons and photons, are promising probes for understanding the various stages of evolution of the fireball created in high energy heavy-ion collisions \cite{Alam:1996fd, Alam:1999sc}. These particles are produced during the entire evolution of the fireball. They interact very weakly with the surrounding matter and escape to the detector without much interaction. Therefore the spectra of thermal particles contain useful information about the evolution of the fireball and should be the most sensitive probe for hydrodynamic attractor. Moreover, thermal particle production from the fireball is found to be sensitive to dissipation and other non-equilibrium effects~\cite{Bhatt:2009zg,Bhatt:2010cy, Chandra:2016dwy}. In this paper, we consider the production of thermal photons and dileptons using Chapman-Enskog type shear viscous corrections in distribution functions~\cite{Bhalerao:2013pza} and study the spectra of these particles for different initial conditions as well a s those corresponding to the attractor solution. We note that, the thermal particle production is studied for the first time employing this form of viscous correction.

The paper is organized as follows as follows: In Section~\ref{hydro}, we present the various causal hydrodynamic formulations used to describe the evolution of the system and their analytical solutions in case of one dimensional longitudinal Bj\"orken expansion. In Section~\ref{alpha_values}, we compare the approximate analytical attractors with exact numerical attractor to quantify the accuracy of the analytical solutions. Further, based on argument of positivity of energy density, we constrain the allowed region in the basin of attraction. Sections~\ref{particle_prod} and~\ref{SPECTRA}  are devoted to calculations of the non-equilibrium thermal dilepton and photon rates. In Section~\ref{results}, we compute thermal particle spectra from the evolving QGP for different initial conditions and compare it with the spectra obtained using the attractor solutions. In Section~\ref{summary}, we summarize our findings with future outlook.

\section{Dissipative hydrodynamics and Bjorken flow}
\label{hydro}

We consider a system with conformal symmetry which implies that the equation of state relating the energy density, $\epsilon$, and pressure, $P$, takes the form $\epsilon=3P$. The energy-momentum tensor for such a system can be written in the Landau frame as
\begin{align}\label{tmunu}
T^{\mu\nu} = \epsilon u^\mu u^\nu - P\Delta ^{\mu \nu} +  \pi^{\mu\nu},  
\end{align}
where $u^\mu$ is the fluid four-velocity and $\Delta^{\mu\nu} \equiv g^{\mu\nu}{-}u^{\mu}u^{\nu}$ is a projection operator orthogonal to $u^{\mu}$. For a conformal system, bulk viscosity vanishes and dissipation in the energy-momentum conservation is only due to shear stress tensor, $\pi^{\mu\nu}$. The metric convention used in this article is $g^{\mu\nu} = {\rm diag}(+\,-\,-\,-)$.

Energy-momentum conservation, i.e., $\partial_\mu T^{\mu\nu}=0$, along and orthogonal to fluid four-velocity leads to evolution equations for $\epsilon$ and $u^\mu$,
\begin{align}
\dot\epsilon + (\epsilon+P)\theta - \pi^{\mu\nu}\sigma_{\mu\nu} &= 0, \label{evol01}\\
(\epsilon{+}P)\,\dot u^\alpha - \nabla^\alpha P + \Delta^\alpha_\nu \partial_\mu \pi^{\mu\nu}  &= 0, \label{evol02}
\end{align}
were we have used the notations $\dot A\equiv u^\mu \partial_\mu A$ for co-moving derivatives, $\theta\equiv \partial_\mu u^\mu$ for the expansion scalar, $\nabla^\alpha\equiv\Delta^{\mu\alpha} \partial_\mu$ for space-like derivatives, and $\sigma_{\mu\nu}\equiv \frac{1}{2} (\nabla_{\mu}u_{\nu}{+}\nabla_{\nu}u_{\mu}) - \frac{1}{3} \theta \Delta_{\mu\nu}$ for the velocity shear tensor.

In order to close the set of hydrodynamic evolution equations, \eqref{evol01} and \eqref{evol02}, we need additional equation for the shear stress tensor $\pi^{\mu\nu}$. The relativistic Navier-Stokes form is the simplest expression of $\pi^{\mu\nu}$ which is obtained at first order in gradients of velocity, $\pi^{\mu\nu} = 2\eta\sigma^{\mu\nu}$, where $\eta$ is the coefficient of shear viscosity. However, relativistic Navier-Stokes theory in Landau frame is unstable and violates causality%
	\footnote{ Recently, there has been some interesting developments in the formulation of causal and stable first-order theories of relativistic dissipative fluid dynamics \cite{Bemfica:2019knx, Kovtun:2019hdm, Das:2020fnr, Hoult:2020eho, Bemfica:2020zjp}}. 
One way to restore causality is to consider higher-order gradient corrections to relativistic Navier-Stokes expression which leads to a relaxation-type equation for $\pi^{\mu \nu}$. This was first proposed by M\"uller, Israel and Stewart \cite{Israel:1979wp, Muller:1967zza, Israel:1976tn}. We consider the minimal causal theory with conformal symmetry \cite{Baier:2006um, Baier:2007ix} and refer to it as the ``MIS'' theory:
\begin{equation}\label{MIS}
   \tau_\pi\dot\pi^{\langle\mu\nu\rangle} + \pi^{\mu\nu} 
   = 2\eta\sigma^{\mu\nu} -\frac{4}{3}\tau_\pi\pi^{\mu\nu}\theta.
\end{equation}
Here $\tau_{\pi}$ is the shear relaxation time and $\dot{\pi}^{\langle\mu\nu\rangle} = \Delta^{\mu\nu}_{\alpha\beta} \dot{\pi}^{\alpha\beta}$ with $\Delta^{\mu\nu}_{\alpha\beta}\equiv\frac{1}{2}(\Delta^\mu_\alpha\Delta^\nu_\beta{+}\Delta^\mu_\beta\Delta^\nu_\alpha) - \frac{1}{3}\Delta^{\mu\nu}\Delta_{\alpha\beta}$ being the projection operator which projects the traceless and symmetric part of a two rank tensor.

Within the framework of relativistic kinetic theory, a systematic formulation of second-order (``transient'') relativistic hydrodynamics was performed in Ref.~\cite{Denicol:2012cn}. Using 14-moment approximation and relaxation-time approximation for the collision term, the evolution equation for $\pi^{\mu\nu}$ for a system of massless particles (conformal system) takes the form:
\begin{equation}\label{DNMR}
\dot{\pi}^{\langle\mu\nu\rangle} \!+ \frac{\pi^{\mu\nu}}{\tau_\pi}\!= 
2\beta_{\pi}\sigma^{\mu\nu}
\!+2\pi_\gamma^{\langle\mu}\omega^{\nu\rangle\gamma}
\!-\frac{10}{7}\pi_\gamma^{\langle\mu}\sigma^{\nu\rangle\gamma} 
\!-\frac{4}{3}\pi^{\mu\nu}\theta.
\end{equation}
Here $\beta_\pi\equiv\eta/\tau_\pi = 4P/5$, and  $\omega^{\mu\nu}\equiv\frac{1}{2}(\nabla^\mu u^\nu{-}\nabla^\nu u^\mu)$ is the vorticity tensor. In the following, we refer to the above equation as ``DNMR'' theory \cite{Denicol:2012cn} which can also be obtained using a Chapman-Enskog like iterative solution of the Boltzmann equation in the relaxation-time approximation~\cite{Jaiswal:2013npa}. 

Extending the Chapman-Enskog like iterative solution to one higher-order, a third-order evolution equation for $\pi^{\mu\nu}$ was derived in Ref.~\cite{Jaiswal:2013vta}:
\begin{align}\label{TOSHEAR}
\dot{\pi}^{\langle\mu\nu\rangle} =& -\frac{\pi^{\mu\nu}}{\tau_\pi}
+2\beta_\pi\sigma^{\mu\nu}
+2\pi_{\gamma}^{\langle\mu}\omega^{\nu\rangle\gamma}
-\frac{10}{7}\pi_\gamma^{\langle\mu}\sigma^{\nu\rangle\gamma}  \nonumber \\
&-\frac{4}{3}\pi^{\mu\nu}\theta
+\frac{25}{7\beta_\pi}\pi^{\rho\langle\mu}\omega^{\nu\rangle\gamma}\pi_{\rho\gamma}
-\frac{1}{3\beta_\pi}\pi_\gamma^{\langle\mu}\pi^{\nu\rangle\gamma}\theta \nonumber \\
&-\frac{38}{245\beta_\pi}\pi^{\mu\nu}\pi^{\rho\gamma}\sigma_{\rho\gamma}
-\frac{22}{49\beta_\pi}\pi^{\rho\langle\mu}\pi^{\nu\rangle\gamma}\sigma_{\rho\gamma} \nonumber \\
&-\frac{24}{35}\nabla^{\langle\mu}\left(\pi^{\nu\rangle\gamma}\dot u_\gamma\tau_\pi\right)
+\frac{4}{35}\nabla^{\langle\mu}\left(\tau_\pi\nabla_\gamma\pi^{\nu\rangle\gamma}\right) \nonumber \\
&-\frac{2}{7}\nabla_{\gamma}\left(\tau_\pi\nabla^{\langle\mu}\pi^{\nu\rangle\gamma}\right)
+\frac{12}{7}\nabla_{\gamma}\left(\tau_\pi\dot u^{\langle\mu}\pi^{\nu\rangle\gamma}\right) \nonumber \\
&-\frac{1}{7}\nabla_{\gamma}\left(\tau_\pi\nabla^{\gamma}\pi^{\langle\mu\nu\rangle}\right)
+\frac{6}{7}\nabla_{\gamma}\left(\tau_\pi\dot u^{\gamma}\pi^{\langle\mu\nu\rangle}\right) \nonumber \\
&-\frac{2}{7}\tau_\pi\omega^{\rho\langle\mu}\omega^{\nu\rangle\gamma}\pi_{\rho\gamma}
-\frac{2}{7}\tau_\pi\pi^{\rho\langle\mu}\omega^{\nu\rangle\gamma}\omega_{\rho\gamma} \nonumber \\
&-\frac{10}{63}\tau_\pi\pi^{\mu\nu}\theta^2
+\frac{26}{21}\tau_\pi\pi_\gamma^{\langle\mu}\omega^{\nu\rangle\gamma}\theta.
\end{align}
In the following, we refer to the above equation as the ``third-order" theory. We mention here that a complete analysis of all possible terms at third-order in gradients was performed in Ref.~\cite{Grozdanov:2015kqa}. The specific form in the above equation reflects transport coefficients obtained from the Boltzmann equation in the relaxation-time approximation~\cite{Jaiswal:2013vta}. Next, we consider the solutions of the three variants of causal dissipative hydrodynamic theories, ``MIS", ``DNMR" and ``third-order" as discussed above, in the simple but physically relevant case of one-dimensional longitudinal Bjorken expansion. 


\subsection*{Bjorken Flow}
\label{bjorken}

In heavy-ion collisions at ultra-relativistic energies, the colliding nuclei approach each other approximately along light-cone trajectories. In order to consider such a system, the Milne coordinates $x^\mu=(\tau,r,\varphi,\eta_s)$ are the natural choice where, $\tau=\sqrt{t^2{-}z^2}$, $r=\sqrt{x^2+y^2}$, $\varphi={\rm atan2}(y,x)$ and $\eta_s=\tanh^{-1}(z/t)$. For longitudinal boost-invariant Bjorken expansion~\cite{Bjorken:1982qr} of a transversely homogeneous system, the hydrodynamic quantities such as the energy density, $\epsilon$, pressure, $P$, and shear stress, $\pi^{\mu\nu}$, depends only on the longitudinal proper time $\tau$. In this scenario, the energy density evolution, Eq.~\eqref{evol01}, and shear evolution, Eqs.~\eqref{MIS}-\eqref{TOSHEAR} can be written in following generic form~\cite{Jaiswal:2019cju}
\begin{align}
  \frac{d\epsilon}{d\tau} &= -\frac{1}{\tau}\left(\frac{4}{3}\epsilon -\pi\right), 
\label{bde1}\\
  \frac{d\pi}{d\tau} &= - \frac{\pi}{\tau_\pi} + \frac{1}{\tau}\left[\frac{4}{3}\beta_\pi 
  - \left( \lambda + \frac{4}{3} \right) \pi - \chi\frac{\pi^2}{\beta_\pi}\right],
  \label{bde2}
\end{align}
where $\pi \equiv -\tau^2 \pi^{\eta_s\eta_s}$. The coefficients $\beta_\pi$, $\lambda$, and $\chi$ appearing in the above equation are given in Table~\ref{coeff} for the three different causal theories studied in this work.

\begin{table}[t!]
 \begin{center}
  \begin{tabular}{|c|c|c|c|c|c|}
   \hline
   & $\beta_\pi$ & $a$ & $\lambda$ & $\chi$ & $\gamma$ \\
   \hline
   MIS & $4P/5$ & $4/15$ & $0$ & $0$ & $4/3$ \\
   \hline
   DNMR & $4P/5$ & $4/15$ & $10/21$ & $0$ & $4/3$\\
   \hline
   ~Third-order~ & $~4P/5~$ & $~4/15~$ & $~10/21~$ & $~72/245~$ & $~412/147~$ \\
   \hline
  \end{tabular}
  \caption{Coefficients appearing in Bjorken flow evolution equation of shear stress tensor, Eqs.~\eqref{bde2} and \eqref{rbde2}, for the three theories considered in this work.}
  \label{coeff}
 \end{center}
 \vspace*{-6mm}
\end{table}

Noting that $\beta_\pi=4P/5=4\epsilon/15$ for all cases, it is convenient to rewrite Eqs.~\eqref{bde1} and \eqref{bde2} in the form
\begin{align}
   &\frac{1}{\epsilon {\tau}^{4/3}} \frac{d(\epsilon {\tau}^{4/3})}{d \tau} = \frac{4}{3}\frac{\bar{\pi}}{\tau}, 
\label{rbde1} \\
   &\frac{d\bar{\pi}}{d \tau } =  - \frac{\bar{\pi}}{{\tau}_{\pi}} + \frac{1}{\tau} 
      \left( a - \lambda \bar{\pi} - \gamma {\bar{\pi}^2} \right). 
\label{rbde2} 
\end{align} 
where, $\bar\pi\equiv\pi/(\epsilon{+}P)=\pi/(4P)$ is the normalized shear stress tensor, or equivalently, the inverse Reynolds number. 
The coefficients $a$, $\lambda$ and $\gamma$ appearing in the above equation are given in Table~\ref{coeff} for the three different 
theories under consideration. 

\begin{table}[t!]
 \begin{center}
  \begin{tabular}{|c|c|c|c|c|}
   \hline 
   \phantom{$\big|$} \!\!\!$T(\tau)$\!\!\! \phantom{$\big|$}
   & $w$ & $\Lambda$ & $k$ & $m$ \\
   \hline
   \phantom{$\Big|$} \!\!\!const.\!\!\!  \phantom{$\Big|$}
   & $\bar{\tau}$ & $ -1  $ & $-\frac{1}{2} \left(\lambda{+}1\right)$ 
   & $\frac{1}{2} \sqrt{4a\gamma{+}\lambda^2}$  \\
   \hline
   \phantom{$\Big|$} \!\!\!ideal\!\!\!  \phantom{$\Big|$}
   & $\frac{3}{2}\bar{\tau}$ & $-\frac{3}{2}$ & $-\frac{1}{4}\left(3\lambda{+}2 \right) $ 
   & $\frac{3}{4} \sqrt{ 4 a \gamma{+}\lambda^2  }$  \\
   \hline
   \phantom{$\Big|$} \!\!\!NS\!\!\!  \phantom{$\Big|$}
   & $\,\frac{3}{2}\!\left( \bar{\tau}{+}\frac{a}{2} \right)\,$ & $\,-\frac{3}{2}\,$ 
   & $\,-\frac{1}{4}\!\left[3\!\left(\lambda{-}\frac{a}{2}\right)\!{+}2 \right]\,$ & $\,\frac{3}{4} \sqrt{ 4 a \gamma{+}\!\left(\lambda{-}\frac{a}{2}\right)^2 }\,$  
   \\
   \hline
  \end{tabular}
  \caption{Arguments and parameters of Eqs.~\eqref{generic_pibar} and \eqref{generic_energy} for the analytic approximate solutions.
  \label{T2}}
 \end{center}
 \vspace*{-6mm}
\end{table}

It is possible to solve the above set of coupled ordinary differential equations analytically with certain approximations for the relaxation time $\tau_\pi$ \cite{Denicol:2017lxn, Jaiswal:2019cju, Denicol:2019lio}. For a conformal system, $\epsilon{\,\propto\,}T^4$ and also from kinetic theory, one has $T\tau_{\pi} = 5\left(\eta/s\right)=\mathrm{const.}$, where  $s$ is the entropy density. The approximate analytical solutions obtained for normalized shear stress and energy density in Ref.~\cite{Jaiswal:2019cju} for three cases (for $\tau_\pi\propto 1/T$ where $T$ is either constant or has proper-time evolution following ideal or Navier-Stokes hydrodynamic solutions) can be expressed in the generic form 
\begin{align}
\bar{\pi}(\bar{\tau})= & \frac{(k{+}m{+}\frac{1}{2}) M_{k+1,m}(w) - \alpha \, W_{k+1,m}(w)}{\gamma |\Lambda| \left[ M_{k,m}(w)+\alpha \, W_{k,m}(w) \right]},\label{generic_pibar}\\ 
\epsilon(\bar{\tau})= & \epsilon_0 \left(\frac{w_{0}}{w}\right)^{\!\frac{4}{3} \left(|\Lambda|-\frac{k}{\gamma}\right)}e^{-\frac{2 }{3 \gamma} \left( w-{w_0} \right)}\nonumber\\ 
&\times \left(\frac{M_{k,m}(w) + \alpha \, W_{k,m}(w) }{M_{k,m}(w_0) + \alpha \, W_{k,m}(w_0)}\right)^{\frac{4}{3 \gamma}}, \label{generic_energy}
\end{align}
where, $\bar{\tau}\equiv\tau/\tau_\pi$ is the scaled proper-time variable (inverse Knudsen number for Bjorken flow), and $M_{k,m}(w)$ and $W_{k,m}(w)$ are Whittaker functions. Here $\epsilon_0$ is the initial energy density at time ${\bar{\tau}}_0$, and $\alpha$ is the integration constant which encodes the initial normalized shear stress $\bar\pi_0$ and initial energy density $\epsilon_0$. The arguments and parameters of Whittaker functions appearing in the above solution for the three cases are given in Table~\ref{T2}. 

In order to obtain the corresponding exact numerical solutions for $\epsilon$ and $\bar{\pi}$, it is convenient to express Eqs.~\eqref{rbde1} and \eqref{rbde2} in the form,
\begin{align}
\frac{1}{\epsilon} \dfrac{d\epsilon}{d\bar{\tau}} &= 
-\frac{4}{\bar{\tau}} \left( \frac{1-\bar{\pi}}{\bar{\pi}+2}\right) 
\label{deps}\\
\left( \frac{ \bar{\pi} + 2 }{3} \right) \frac{d \bar{\pi}}{d \bar{\tau}} 
&= - \bar{\pi} + \frac{1}{\bar{\tau}} \left( a - \lambda \, \bar{\pi} - \gamma \, \bar{\pi}^2 \right),  \label{dpibar}
\end{align}
where the coefficients $a$, $\lambda$ and $\gamma$ are given in Table~\ref{coeff} for the three different theories considered in this work. In the following, we study the approximate analytical solutions given in Eqs.~\eqref{generic_energy} and \eqref{generic_pibar}, and compare the attractors of MIS, DNMR and third-order theories obtained from these analytical solutions with the corresponding exact numerical attractors.


\section{Attractors and physically allowed region}
\label{alpha_values}

In order to identify the attractor solution using the approximate analytical solution given in Eq.~\eqref{generic_pibar} for MIS, DNMR and third-order theories, we need to find the corresponding value of the constant $\alpha$ corresponding to attractor initial condition. Moreover, it is of importance to check how the approximate analytical attractor compares with the exact numerical attractor solution in these three cases in order to estimate the accuracy of the analytical solutions. Furthermore, it is also essential to constrain the allowed values of the constant $\alpha$ on physical grounds so that one can employ the analytical solutions to calculate spectra of thermal particles. In this section, we address these issues.

\subsection{Values of $\alpha$ at fixed points}
\label{alpha_0}

From Eq.~(\ref{generic_pibar}), we obtain the constant $\alpha$ at the fixed points by imposing the condition,
\begin{align}
   \bar{\pi}_{\pm}=
   \left. \frac{(k{+}m{+}\frac{1}{2}) M_{k+1,m}(w) - \alpha_\pm \, W_{k+1,m}(w)}
             {\gamma |\Lambda| \left[ M_{k,m}(w)+\alpha_\pm \, W_{k,m}(w) \right]} \right|_{w \to 0}.
 \end{align}
where $\bar{\pi}_\pm \equiv \frac{ -\lambda \pm \sqrt{4 a \gamma + \lambda^2}}{2 \gamma} = \frac{1}{\Lambda \gamma} \left( \pm m +k + \frac{1}{2} \right)$  are the attractor and repulsor fixed points corresponding to `$+$' and `$-$' respectively. Solving for $\alpha_\pm$, we obtain,
\begin{align}\label{alpha_fp}
   \alpha_\pm=
   & \left. \frac{(k{+}m{+}\frac{1}{2}) M_{k+1,m}(w) - \bar{\pi}_\pm \,  \gamma \, |\Lambda| \, M_{k,m}(w)}
             {W_{k+1,m}(w) + \bar{\pi}_\pm \,  \gamma \, |\Lambda| \, W_{k,m}(w)} \right|_{w \to 0}.
 \end{align}
In the limit  $w \to 0$, we find $\alpha_{+} =0$ and $\alpha_{-} =\infty$.

The procedure for identifying the hydrodynamic attractor as given in Ref.~\cite{Jaiswal:2019cju} was to look for the value $\alpha_0$ at which the quantity
\begin{equation}\label{psi}
   \psi(\alpha_0) \equiv \lim_{\bar{\tau}  \to 0} 
   \frac{\partial\bar{\pi}}{\partial \alpha}\bigg{|}_{\alpha=\alpha_0}
\end{equation}
diverges. From Eq.~(\ref{generic_pibar}), this was shown to hold for  $\alpha{\,=\,}0$%
	\footnote{Except for the NS approximate solution, for reasons  discussed in Ref.~\cite{Jaiswal:2019cju}.}, which is $\alpha_{+}$ as obtained in the previous section. The attractor solution is given as
\begin{align}\label{gen_att}
   \bar{\pi}_\mathrm{attr}(w)=
   & \frac{k{+}m{+}\frac{1}{2}} {\gamma |\Lambda|}\, \frac{M_{k+1,m}(w)}{M_{k,m}(w)}.
 \end{align}
Different initializations of the system in out of equilibrium regime first converges on this hydrodynamic attractor and then evolves to attain local thermal equilibrium. 

From Eq.~\eqref{psi}, we can also obtain the equation for repulsor which is the other separatrix. We look at $\bar{\tau} = \infty$ slice and look for the value of $\alpha_0$ for which $\psi$ diverges.%
	\footnote{The motivation for $\bar{\tau} = \infty$ choice is from Fig~\ref{stream_plot}(b). As $\bar{\tau} \to \infty$, all curves converge on the attractor, except the repulsor curve.} 
We find $\psi$ diverges for $\alpha = \infty$ (which is $\alpha_{-}$ as obtained in the previous section) and is $0$ for all other $\alpha$.  So the repulsor curve is given by replacing $\alpha$ with $\infty$, which is the same as dropping the first terms in the numerator and denominator of Eq.~\eqref{generic_pibar}:
\begin{align}
\label{repulsor}
   \bar{\pi}_\mathrm{repl}(\bar{\tau})=   - \frac{1}{\gamma |\Lambda|} \frac{ W_{k+1,m}(w)}{ W_{k,m}(w)}.
\end{align}
 
In the following subsections, we compare the analytically obtained attractor with the exact numerical attractor, and determine the physically allowed region of the ``configuration space" and their relations to the attractor and repulsor.

\subsection{Analytical solutions vs numerical solutions}
\label{comparison}
The approximate solutions obtained provides more analytical handle to study particle production. Before discussing thermal particle produced in heavy ion collisions, we first discuss the accuracy of obtained analytical solutions by comparing it with the numerical solution of exact equation~\eqref{dpibar}. Since an early time attractor exist for a conformal fluid undergoing Bjorken expansion, we compare the analytical attractors of various theories with exact numerical attractor.

\begin{figure}[t!]
 \begin{center}
  \includegraphics[width=\linewidth]{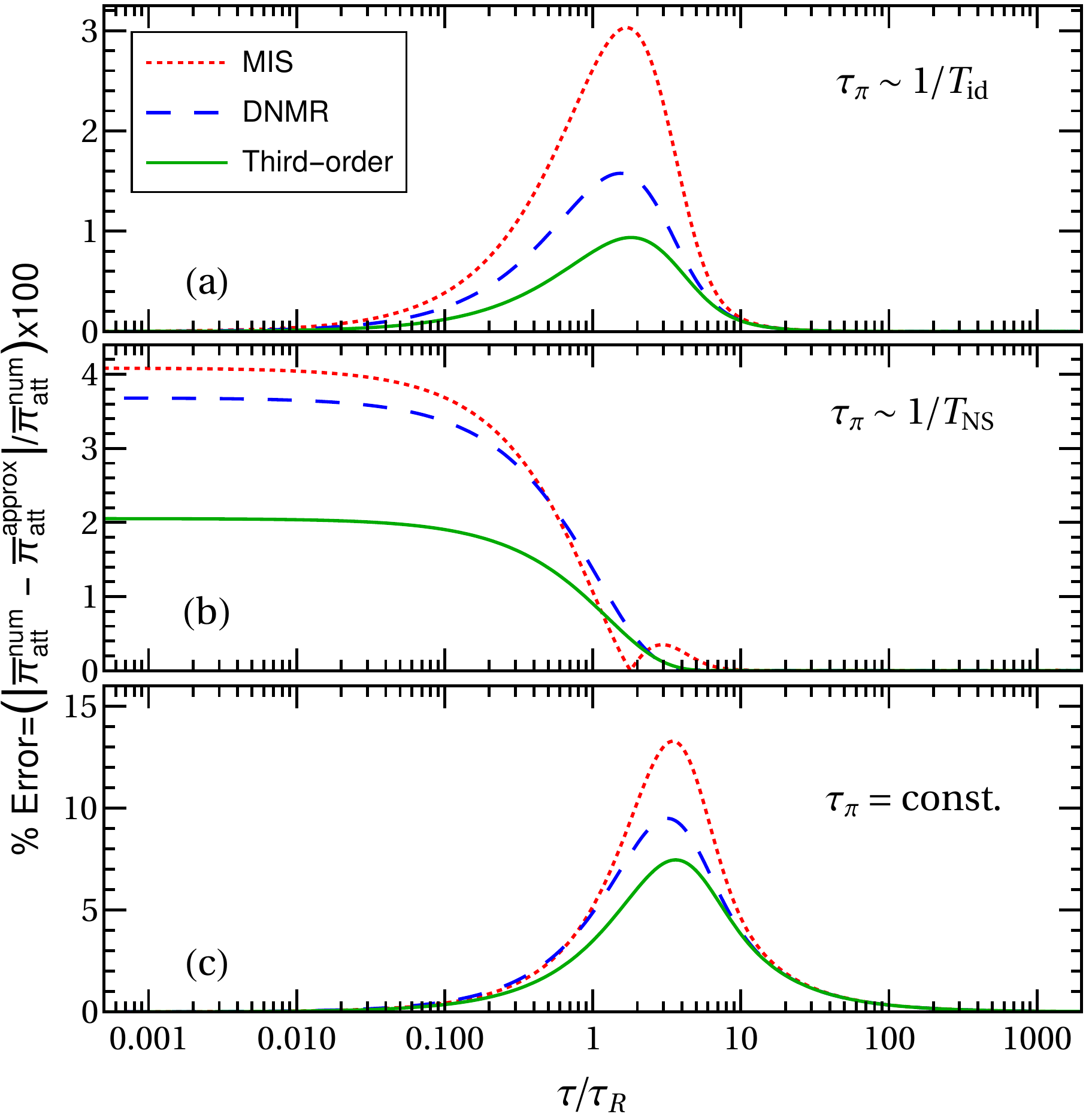}
 \end{center}
 \vspace{-0.6cm}
 \caption{Comparison of exact (numerical) and approximate analytical attractors of different theories.}
 \label{att_comp}
\end{figure}

In Fig.~\ref{att_comp}, we show the the percentage error between the numerically obtained attractor and the  corresponding approximate attractor obtained in Eq.~\eqref{gen_att} for the theories considered. The numerical attractor is obtained by initializing $\bar\pi$ at the stable fixed points of the respective theories and at $\bar\tau  \approx 0 $. First, we note that the approximate solution obtained for the case $\tau_\pi \sim 1/T_{\rm id}$ in Fig.~\ref{att_comp}(a) shows the least error ($\sim 3 \%$) for the corresponding attractors for all theories compared to other approximate attractors, the maximum error occurring during the transition from free streaming to hydrodynamic regime. Further, the solid green curve corresponding to the error for third-order hydrodynamics remains least among the approximate attractors in  all three theories considered. In fact, if one considers the approximate solution obtained using  $\tau_\pi \sim 1/T_{\rm id}$ for third-order hydrodynamics, the error remains $\lesssim 1 \%$ during the entire evolution. This demonstrates that the approximate analytical solutions obtained are in good agreement with exact solutions, and can be used to study thermal particle production.

\subsection{Allowed region in basin of attraction}
\label{basin_att}

We now exploit the analytical solutions to constrain the allowed initial states of the system. An important constraint on the constant of integral $\alpha$ in Eqs.~\eqref{generic_pibar} and \eqref{generic_energy} is that it can only take values for which the energy density is positive-definite for $\bar{\tau}>0$. From Eq.~(\ref{generic_energy}), this implies%
	\footnote{Note that $\frac{M_{k,m}(w)}{W_{k,m}(w)}\geq0$ for all $w$ and $\frac{M_{k,m}(0)}{W_{k,m}(0)}=0$.},
\begin{align}
\label{alpha_allowed} 
& \frac{M_{k,m}(w) + \alpha \, W_{k,m}(w) }{M_{k,m}(w_0) + \alpha \, W_{k,m}(w_0)} \, \geq \, 0, 
\nonumber \\
\implies & \left(\frac{M_{k,m}(w)}{W_{k,m}(w)} + \alpha \right) \Big/ \left(\frac{M_{k,m}(w_0)}{W_{k,m}(w_0)} + \alpha \right) \, \geq \,  0, 
\nonumber \\
\implies & \alpha \geq  \, 0.
\end{align}
This sets the bound on $\alpha \in [0,\infty ]$, the extreme values corresponding to attractor and repulsor solutions.

\begin{figure}[t!]
 \begin{center}
  \includegraphics[width=\linewidth]{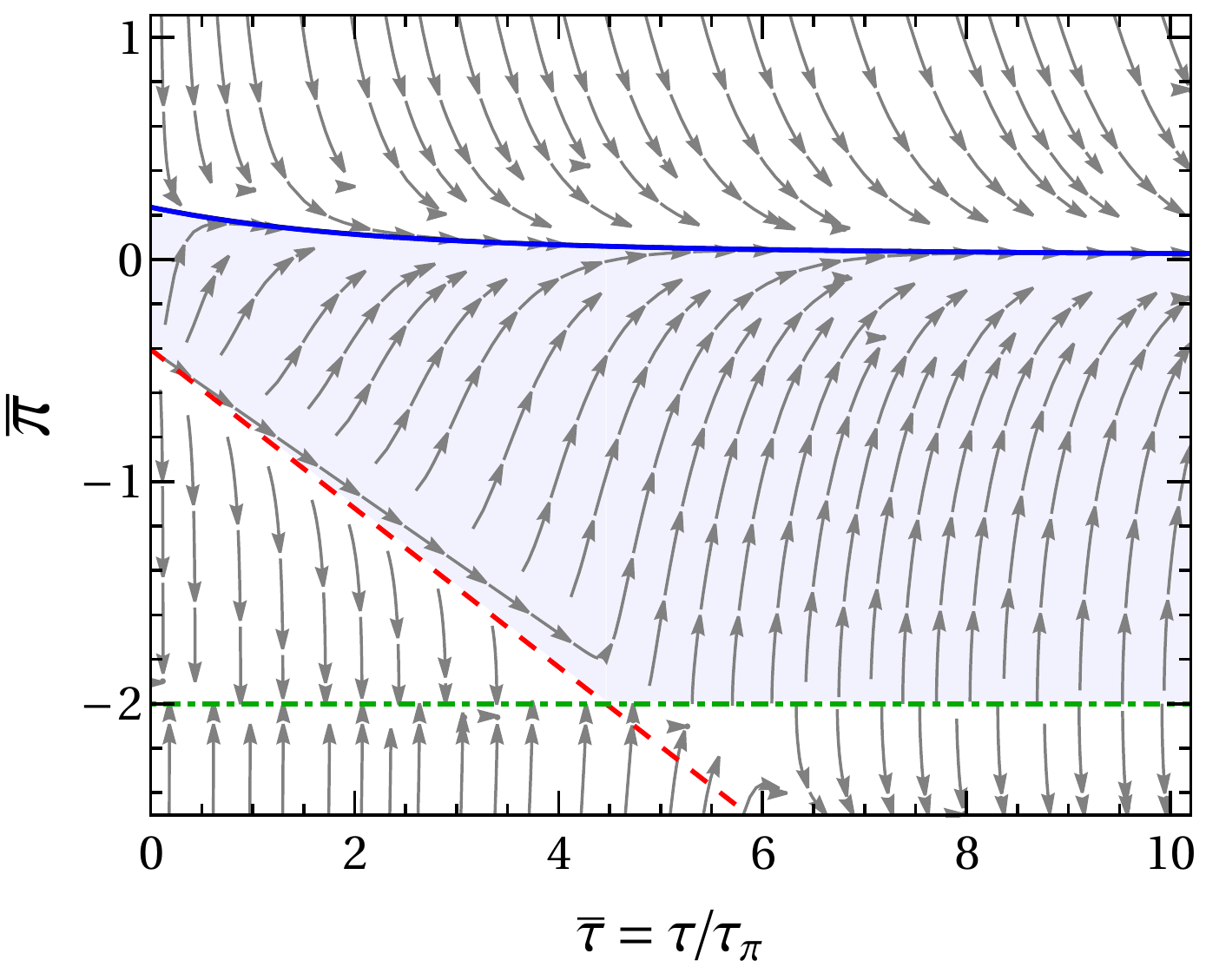}
 \end{center}
 \vspace*{-0.6cm}
 \caption{Evolution of Bjorken configuration space using Eq.~\eqref{dpibar}. Solid blue curve and dashed red curve represents the analytically obtained attractor and repulsor respectively.}
 \label{stream_plot}
  \vspace*{-0.2cm}
\end{figure}

To demonstrate the physically allowed region,  we consider the analytical solutions in the approximation $\tau_\pi \sim 1/T_{\rm id}$ for third order theory since it agrees best with numerical result. Fig.~\ref{stream_plot} represents the evolution trajectories of $\bar{\pi}$ with $\bar{\tau}$ for the exact differential equation~\eqref{dpibar}. The solid blue and dashed red curves represents the analytical attractor~\eqref{gen_att} and repulsor~\eqref{repulsor} respectively. The attractor is seen to be in excellent agreement with the exact numerical attractor as already demonstrated in Fig.~\ref{att_comp}. The repulsor (dashed blue curve) shows some deviation from the numerical repulsor. Positive values of $\alpha$ corresponds to the region between the attractor and the repulsor curves. All other regions in  Fig.~\eqref{stream_plot} corresponds to $-ve$ values of $\alpha$. This restricts our physically allowed region between the separatrix.

The basin of attraction imposes yet another condition on the physically allowed region. The dot-dashed green line in Fig.~\ref{stream_plot} represents a feature of the exact differential equation \eqref{dpibar} which approximate solutions fails to capture. $d \bar{\pi}/d \bar{\tau}$  in Eq.~\eqref{dpibar} diverges at $\bar{\pi} = -2$, and trajectories between the separatrix, but below $\bar{\pi} = -2$ never approaches the attractor. Though this region is in the physically permitted region ($+ve \ \alpha$), we neglect this region as it never approaches the attractor. The blue shaded area in Fig.~\ref{stream_plot} represents the allowed basin of attraction. We note that within this basin of attraction, longitudinal pressure  $P_L = P (1 - 4 \bar{\pi} )$ is always positive and there is no cavitation in the longitudinal direction.


\section{Thermal particle production} 
\label{particle_prod}

Thermal particles, such as dileptons and photons, are produced throughout the evolution of QGP. Dominant mechanism for thermal dilepton production in $\mathrm{QGP}$ arises from $q \bar{q}$ annihilation process, $q \bar{q} \rightarrow \gamma^{*} \rightarrow l^{+} l^{-}$. From kinetic theory, the dilepton production rate for this process can be written as~\cite{Vogt:2007zz}
\begin{align}
\frac{dN_{l^{+}l^{-}}}{d^4x d^4p}=g^2 \int & \frac{d^3\textbf{p}_1}{(2\pi)^3} 
	\frac{d^3\textbf{p}_2}{(2\pi)^3} f(E_1,T) f(E_2,T) 
\nonumber \\
& v_{rel} \sigma(M^2)\delta^4(p-p_1-p_2),
\label{eq1}
\end{align}
where $p_{1,2}=\left(E_{1,2}, \mathbf{p}_{1,2}\right)$ is the four momentum of quark, anti-quark with $E_{1,2}=\sqrt{\mathbf{p}_{1,2}^{2}+m^{2}} \simeq\left|\mathbf{p}_{1,2}\right|$ by taking the quark masses ($m$) to be zero. Here, invariant mass of the virtual photon is represented as $M^{2}=\left(E_{1}+E_{2}\right)^{2}-\left(\mathbf{p}_{1}+\mathbf{p}_{2}\right)^{2}$. The quark (anti-quark) distribution function in thermal equilibrium with the degeneracy factor $g$ is denoted by $f(E, T)$. Relative velocity of the quark-anti-quark pair is denoted as $v_{r e l}=\sqrt{\frac{M^{2}\left(M^{2}-4 m^{2}\right)}{4 E_{1}^{2} E_{2}^{2}}} \sim \frac{M^{2}}{2 E_{1} E_{2}}$ while $\sigma\left(M^{2}\right)$ is the thermal dilepton production cross section for the annihilation process and is given as (with $N_{f}=2$ and $N_{c}=3$ ), $M^{2} g^{2} \sigma\left(M^{2}\right)=\frac{80 \pi}{9} \alpha_{e}^{2}$ ~\cite{Alam:1996fd}, where $\alpha_e$ is the electromagnetic coupling constant.

The effect of viscosity on particle production enters via the expressions for (shear) viscous modified phase-space distribution functions $f=f_0+\delta f$, where $f_{0}$ denotes the ideal part. Here we consider the form of $\delta f$ due to Chapman-Enskog method~\cite{Bhalerao:2013pza}
\begin{equation}
\label{deltaF}
\delta f= \frac{f_0 \beta}{2\beta_{\pi}(u \!\cdot\! p)}\, p^\alpha p^\beta\pi_{\alpha\beta}.
\end{equation}
Now, the viscous modified quark (anti-quark) distribution function can be written as
\begin{equation}
\label{eq2}
f(E, T)=f_{0}(E, T)\left(1+\frac{\beta}{\beta_{\pi}} 
\frac{p^{\alpha} p^{\beta} \pi_{\alpha \beta}}{2(u \cdot p)}\right),
\end{equation}
where, $\beta=1/T$ and $\beta_\pi=(\epsilon+P)/5$. This form of $\delta f$ leads to correct scaling behavior of the longitudinal femtoscopic radii and was shown to be a better alternative compared to popularly used 14-moment ansatz for hydrodynamic 
modeling of relativistic heavy-ion collisions~\cite{Bhalerao:2013pza}.

Substituting Eq.~\eqref{eq2} in dilepton rate i.e., in Eq.~\eqref{eq1} and keeping only linear terms in $\delta f,$ we can write the ideal and viscous contribution to production rate as~\cite{Bhatt:2011kx}
\begin{equation}
 \frac{d N_{l^{+}l^{-}}}{d^{4} x d^{4} p}=\frac{d N_{l^{+}l^{-}}^{0}}{d^{4} x d^{4} p}
 +\frac{d N_{l^{+}l^{-}}^{\pi}}{d^{4} x d^{4} p},
\end{equation}
with
\begin{eqnarray}
\frac{d N_{l^{+}l^{-}}^{0}}{d^{4} x d^{4} p}&=&\int \frac{d^{3} \mathbf{p}_{1}}{(2 \pi)^{3}} \frac{d^{3} \mathbf{p}_{2}}{(2 \pi)^{3}} f_{0}\left(E_{1}, T\right) f_{0}\left(E_{2}, T\right)
\nonumber\\
&&\times\frac{M^{2} g^{2} \sigma\left(M^{2}\right)}{2 E_{1} E_{2}} \delta^{4}\left(p-p_{1}-p_{2}\right) 
\label{eq4}\\
\frac{d N_{l^{+}l^{-}}^{\pi}}{d^{4} x d^{4} p}&=&\int \frac{d^{3} \mathbf{p}_{1}}{(2 \pi)^{3}} \frac{d^{3} \mathbf{p}_{2}}{(2 \pi)^{3}} f_{0}\left(E_{1}, T\right) f_{0}\left(E_{2}, T\right)
\nonumber \\
&&\times\left[\frac{\beta}{\beta_{\pi}(u \cdot p_1)}\right] \frac{M^{2} g^{2} \sigma\left(M^{2}\right)}{2 E_{1} E_{2}} \delta^{4}\left(p-p_{1}-p_{2}\right)
\nonumber \\
&& \times p_1^\alpha\, p_1^\beta\, \pi_{\alpha\beta}.
\end{eqnarray}
Further, the viscous correction part can be written as,
\begin{equation}
\frac{d N_{l^{+}l^{-}}^{\pi}}{d^{4} x d^{4} p}=\frac{\beta}{\beta_{\pi}} 
I^{\alpha \beta} \pi_{\alpha \beta},
\end{equation}
where
\begin{eqnarray}
I^{\alpha \beta}&=&\int \frac{d^{3} \mathbf{p}_{1}}{(2 \pi)^{6}} f_{0}\left(E_{1}, T\right) 
	f_{0}\left(E_{2}, T\right) \frac{p_{1}^{\alpha} p_{1}^{\beta}}{(u \cdot p_1)} 
\nonumber\\
&&\times\frac{M^{2} g^{2} \sigma\left(M^{2}\right)}{2 E_{1} E_{2}} \delta\left(E-E_{1}-E_{2}\right).
\end{eqnarray}

Now we proceed to cast the second rank tensor $I^{\alpha \beta}$ in the most general form using $u^{\alpha}$ and $p^{\alpha}$ :
\begin{equation}
I^{\alpha \beta}=a_{0} g^{\alpha \beta}+a_{1} u^{\alpha} u^{\beta}+a_{2} p^{\alpha} p^{\beta}+a_{3}\left(u^{\alpha} p^{\beta}+u^{\beta} p^{\alpha}\right).
\end{equation}
However, we note that while contracting with $\pi_{\alpha \beta},$ only the coefficient $a_{2}$ remains, since $u^{\alpha} \pi_{\alpha \beta}=0$ and $\pi_{\alpha}^{\alpha}=0 .$ We calculate the surviving coefficient $a_{2}=J_{\alpha \beta} I^{\alpha \beta}$ by constructing the projection operator $J_{\alpha \beta},$ which, in the local rest frame of the medium $\left(u^{\alpha}=(1, \overline{0})\right)$ has the form
\begin{equation*}
J_{\alpha \beta}=\frac{\left[|\mathbf{p}|^{2} g_{\alpha \beta}
+\left(2 E^{2}+M^{2}\right) u_{\alpha} u_{\beta}+3 p_{\alpha} p_{\beta}
-6 E u_{\alpha} p_{\beta}\right]}{2|\mathbf{p}|^{4}}.
\end{equation*}

Viscous correction to the dilepton rate can be calculated in this frame and the final expression is written as
\begin{eqnarray}
\label{dilepton_pro}
\frac{d N_{l^{+}l^{-}}^{\pi}}{d^{4} x d^{4} p}&=&\frac{\beta}{\beta_{\pi}}\left[J_{\alpha \beta} 
I^{\alpha \beta}\right] p^{\alpha} p^{\beta} \pi_{\alpha \beta}
\nonumber\\
&=&\frac{\beta}{\beta_{\pi}} \frac{M^{2} g^{2} \sigma\left(M^{2}\right)}{2(2 \pi)^{5}} 
H_{0}(p) p^{\alpha} p^{\beta} \pi_{\alpha \beta},
\end{eqnarray}
with
\begin{eqnarray}
H_{0}(p)&=&\frac{1}{2|\mathbf{p}|^{5}} \int_{\left|\mathbf{p}_{1}\right|_-}^{\left|\mathbf{p}_{1}\right|_+} 
d\left|\mathbf{p}_{1}\right| f_{0}\left(E_{1}, T\right) f_{0}\left(E-E_{1}, T\right)
\nonumber\\
&&\times\left[\left(2 E^{2}+M^{2}\right)\left|\mathbf{p}_{1}\right|+\frac{3}{4}
\frac{M^{4}}{\left|\mathbf{p}_{1}\right|}-3 E M^{2}\right].
\end{eqnarray}

Here values of limits of integration are $\left|\mathbf{p}_{1}\right|_{\mp}=M^{2} / 2(E \pm|\mathbf{p}|).$ Since we are interested in large invariant mass dileptons $M \geqslant T,$ equilibrium distribution functions can be approximated with Maxwell-Boltzmann expressions:
$f_{0}(E, T) \approx e^{-E / T}$. Under this assumption, ideal part of the dilepton production rate given by Eq.(\ref{eq4}) becomes~\cite{Vogt:2007zz}
\begin{equation}
\frac{d N_{l^{+}l^{-}}^{0}}{d^{4} x d^{4} p}=\frac{1}{2} \frac{M^{2} g^{2} \sigma
\left(M^{2}\right)}{(2 \pi)^{5}} e^{-E / T}.
\end{equation}
Further, by evaluating $H_{0}(p)$ in the Maxwell-Boltzmann limit,
\begin{eqnarray*}
H_{0}(p) &\approx& \frac{e^{-E / T}}{2|\mathbf{p}|^{5}}
\left[\frac{E|\mathbf{p}|}{2}\left(2|\mathbf{p}|^{2}-3 M^{2}\right)
+\frac{3}{4} M^{4} \ln \left(\frac{E+|\mathbf{p}|}{E-|\mathbf{p}|}\right)\right].
\end{eqnarray*}
With this, the contribution of shear viscosity to dilepton production rate is obtained from Eq.~\eqref{dilepton_pro} as
\begin{eqnarray}
\frac{d N_{l^{+}l^{-}}^{\pi}}{d^{4} x d^{4} p}&=&\frac{d N_{l^{+}l^{-}}^{0}}{d^{4} x d^{4} p}
\Bigg\{\frac{\beta}{\beta_{\pi}} \frac{1}{2|\mathbf{p}|^{5}}
\Bigg[\frac{E|\mathbf{p}|}{2}\left(2|\mathbf{p}|^{2}-3 M^{2}\right)
\nonumber\\
&&+\frac{3}{4} M^{4} \ln \left(\frac{E+|\mathbf{p}|}{E-|\mathbf{p}|}\right)\Bigg]
p^{\alpha} p^{\beta} \pi_{\alpha \beta}\Bigg\}.
\end{eqnarray}
We note that this form for production rate is similar to the form obtained in Ref.~\cite{Chandra:2020hik} in which non-equilibrium effects due to chromo-turbulent fields in a collisional non-viscous hot QCD medium were considered.

Similarly, one can calculate the modification to thermal photon rates arising from the inclusion of viscosity. Considering prominent sources of thermal photons, Compton scattering: $q(\bar{q}) g \rightarrow q(\bar{q}) \gamma$ and $q \bar{q}$ -annihilation $q \bar{q} \rightarrow g \gamma$, the final expression for total photon production rate can be written as~\cite{Dusling:2009bc,Bhatt:2010cy,Wong:1995jf}
\begin{align}
\label{PH1}
 E \frac{d N_{\gamma}}{d^{4} x d^{3} p} &= E \frac{d N_{\gamma}^{0}}{d^{4}x d^{3}p}
 +E \frac{d N_{\gamma}^{\pi}}{d^{4}x d^{3}p}
 \nonumber\\
 &=\frac{5}{9} \frac{\alpha_{e} \alpha_{s}}{2 \pi^{2}} f(E, T) T^{2} 
 \left[ \ln \left(\frac{12 E}{g^{2} T}\right)
 \!+\! \frac{C_{\textrm{ann}} \!+\! C_{\textrm{Comp}}}{2} \right],  
\end{align}
where the constants take the values $C_{\textrm{ann}} = -1.91613$, $C_{\textrm{Comp}} = -0.41613$ and $g=\sqrt{4\pi \alpha_s}$; with $\alpha_s$ denoting the strong coupling constant.
 
With the expression for modified distribution function, Eq.~\eqref{eq2} under Maxwell-Boltzmann limit, we can write the ideal and viscous contribution to the rate, respectively, as:
\begin{eqnarray}
E \frac{d N_{\gamma}^{0}}{d^{4} x d^{3} p}&=&\frac{5}{9} \frac{\alpha_{e} \alpha_{s}}{2 \pi^{2}} 
T^{2} e^{-E / T} \ln \left(\frac{3.7388 E}{g^{2} T}\right), \\
E \frac{d N_{\gamma}^{\pi}}{d^{4} x d^{3} p}&=&E \frac{d N_{\gamma}^{0}}{d^{4} x d^{3} p}
\left\{\frac{\beta}{\beta_{\pi}} \frac{p^{\alpha} p^{\beta} \pi_{\alpha \beta}}{2 E}\right\}.
\end{eqnarray}
Finally, we need to cast the above calculated thermal particle production rate expressions in the local rest frame of the medium, 
into a general frame with four-velocity $u^{\mu}$. By noting the relations: 
$E=u \cdot p$ and $\pi_{\alpha \beta}/\beta_{\pi}=5\bar{\pi}_{\alpha \beta}$, the desired expressions are given as
\begin{eqnarray}
\frac{d N_{l^{+}l^{-}}^{0}}{d^{4} x d^{4} p} &=&\frac{1}{2} \frac{M^{2} g^{2} \sigma
\left(M^{2}\right)}{(2 \pi)^{5}} e^{-u \cdot p / T} \,, 
\\
\frac{d N_{l^{+}l^{-}}^{\pi}}{d^{4} x d^{4} p} &=&\frac{d N_{l^{+}l^{-}}^{0}}{d^{4} x d^{4} p}
\Bigg\{\frac{5\beta}{2\left[(u \cdot p)^{2}-M^{2}\right]^{5 / 2}}
\nonumber \\
&& \times\Bigg[\frac{(u \cdot p) \sqrt{(u \cdot p)^{2}-M^{2}}}{2}
\left(2(u \cdot p)^{2}-5 M^{2}\right) 
\nonumber \\
&&+\frac{3}{4} M^{4} \ln \left(\frac{u \cdot p+\sqrt{(u \cdot p)^{2}-M^{2}}}
{u \cdot p-\sqrt{(u \cdot p)^{2}-M^{2}}}\right)\Bigg]\Bigg\} 
\nonumber \\
&& \times \, p^{\alpha} p^{\beta} \bar{\pi}_{\alpha \beta} \,, \\
E \frac{d N_{\gamma}^{0}}{d^{4} x d^{3} p} &=&\frac{5}{9} \frac{\alpha_{e} \alpha_{s}}{2 \pi^{2}} 
T^{2} e^{-u \cdot p / T} \ln \left(\frac{3.7388(u \cdot p)}{g^{2} T}\right) \,, \\
E \frac{d N_{\gamma}^{\pi}}{d^{4} x d^{3} p} &=&E \frac{d N_{\gamma}^{0}}{d^{4} x d^{3} p}
\left\{ \frac{5\beta}{2(u \cdot p)}\right\}p^{\alpha} p^{\beta} \bar{\pi}_{\alpha \beta} \,.
\end{eqnarray}
After incorporation of viscous effects in the thermal particle rates, we next turn our attention to thermal spectra produced during hydrodynamical evolution of the system.

\vspace*{-0.2cm}
\section{Thermal particle spectra from heavy ion collision}
\label{SPECTRA}
\vspace*{-.2cm}

Total thermal dilepton (photon) spectrum can be obtained by convoluting dilepton (photon) production rate with the space-time evolution of QGP in relativistic heavy-ion collision. For Bjorken model, the four dimensional volume element is given by $d^{4} x=\pi R_{A}^{2} d \eta_{s} \tau d \tau$, with $R_{A}=1.2 A^{1 / 3}$ being the radius of the colliding nuclei ($A=197$ for Au). We now calculate the thermal particle (dilepton and photon) yields in terms of their invariant mass $(M)$, transverse momenta $\left(p_{T}\right)$, and rapidity ($y$) as:
\begin{align}
\frac{d N_{l^{+}l^{-}}}{d M^2 d^2 p_{T} d y} &= \pi R_{A}^{2} 
\int_{\tau_{0}}^{\tau_{f}} d \tau\ \tau \int_{-\infty}^{\infty} d \eta_{s}
\left(\frac{1}{2} \frac{d N_{l+l-}}{d^{4} x d^{4} p}\right), \\
\frac{d N_{\gamma}}{d^{2} p_{T} d y} &= \pi R_{A}^{2} \int_{\tau_{0}}^{\tau_{f}} d \tau\ \tau 
\int_{-\infty}^{\infty} d \eta_{s}\left(E \frac{d N_{\gamma}}{d^{3} p d^{4} x}\right).
\end{align}
Here $\tau_{0}$ and $\tau_{f}$ are the initial and final values of the system evolution time that we are interested.

In $(\tau,r,\varphi,\eta_s)$ coordinates, the components of particle four-momenta are given by
\begin{align}
p^\tau &= m_T \cosh(y-\eta_s), \quad p^r = p_T \cos(\varphi_p- \varphi), 
\\
p^\varphi &= p_T\sin(\varphi_p-\varphi)/r, \quad p^{\eta_s} = m_T \sinh(y-\eta_s)/\tau, 
\nonumber
\end{align}
where $m_T^2=p_T^2+M^2$. Here $p_T$ is the transverse momentum, $y$ is the particle rapidity, and $\varphi_p$ is the azimuthal angle in the momentum space. Now, for the expanding medium under Bjorken flow, the factors appearing in the rate expressions are obtained as 
$u\cdot p=$ $m_{T} \cosh \left(y-\eta_{s}\right)$ and
\begin{equation}
p^{\alpha} p^{\beta} \bar{\pi}_{\alpha \beta}=\bar{\pi}\left[\frac{p_{T}^{2}}{2}-m_{T}^{2} \sinh^{2}
\left(y-\eta_{s}\right)\right],
\end{equation}
where $\pi\equiv -\tau^2 \pi^{\eta\eta}$ in Milne coordinates.

The ideal contribution to thermal dilepton yield can be given as
\begin{equation}\label{dilid-yield}
 \frac{d N_{l^{+}l^{-}}^{0}}{d M^2 d^2 p_{T} d y} = 2 \mathscr{F} 
 \int_{\tau_0}^{\tau_f}d\tau\,\tau K_0(z_T), 
\end{equation}
where $K_n$ is the modified Bessel function of second kind, $z_T\equiv m_T/T$, and $\displaystyle{\mathscr{F}=\frac{R_A^2}{2^5 \pi^3}\frac{20}{9}\alpha_e^2}$. 

The viscous contribution to the thermal dilepton yield is obtained as follows:
\begin{equation}
\frac{d N_{l^{+}l^{-}}^{\pi}}{d M^2 d^2 p_{T} d y} = \mathscr{F}\!\!\!\! 
\int_{\tau_0}^{\tau_f}\!\!d\tau \tau\int_{-\infty}^{\infty}\!\!\!\!d\eta_s\, 
e^{-z_T\cosh(y-\eta_s)} \mathscr{E}(T,\eta_s),
\end{equation}
where
\begin{widetext}
\begin{eqnarray}
 \mathscr{E}(T,\eta_s) =   \frac{5\bar{\pi}}{2T}\frac{\left[p_T^2/2-m_T^2\sinh^2(y-\eta_s)\right]}{[p_T^2+m_T^2\sinh^2(y-\eta_s)]^{5/2}}
 &&\Bigg[\left(2m_T^2\cosh^2(y-\eta_s)-5M^2\right)\sqrt{p_T^2+m_T^2\sinh^2(y-\eta_s)}
 \frac{m_T\cosh(y-\eta_s)}{2} \nonumber\\
 &&\ +\frac{3}{4}M^4\ln\left(\frac{m_T\cosh(y-\eta_s)
 +\sqrt{p_T^2+m_T^2\sinh^2(y-\eta_s)}}{m_T\cosh(y-\eta_s)
 -\sqrt{p_T^2+m_T^2\sinh^2(y-\eta_s)}}\right)\Bigg]\,.
\end{eqnarray}
\end{widetext}

Now, the total thermal dilepton yield can be written as sum of ideal and viscous contributions:
\begin{eqnarray}
 \frac{d N_{l^{+}l^{-}}}{d M^2 d^2 p_{T} d y} = \frac{d N_{l^{+}l^{-}}^{0}}{d M^2 d^2 p_{T} d y}
 +\frac{d N_{l^{+}l^{-}}^{\pi}}{d M^2 d^2 p_{T} d y}.
\end{eqnarray}
Similarly, the total photon yield is given by,
\begin{equation}
 \frac{d N_{\gamma}}{d^{2} p_{T} d y} = \frac{d N_{\gamma}^{0}}{d^{2} p_{T} d y} 
 + \frac{d N_{\gamma}^{\pi}}{d^{2} p_{T} d y}.
\end{equation}
Noting the photon energy to be $p_T\cosh(y-\eta_s)$, we write the ideal part of 
thermal photon yield as
\begin{eqnarray}
\label{phID-yield}
 \frac{d N_{\gamma}^{0}}{d^{2} p_{T} d y} &=& \mathscr{G} \int_{\tau_{0}}^{\tau_{f}} d \tau  
 T^2 \tau \int_{-\infty}^{\infty}d\eta_s e^{-p_T/T\cosh(y-\eta_s)} 
 \nonumber\\
 &&\times\ln \left(\frac{3.7388\,p_T\cosh(y-\eta_s)}{g^{2} T}\right) ,
\end{eqnarray}
where $\mathscr{G}=\pi R_A^2 \frac{5}{9} \frac{\alpha_{e} \alpha_{s}}{2 \pi^{2}}$.

The viscous contribution to the photon spectra is given by,
\begin{equation}
  \frac{d N_{\gamma}^{\pi}}{d^{2} p_{T} d y} = \mathscr{G}\!\! \int_{\tau_{0}}^{\tau_{f}} \!\!\!\! d \tau\,  
  T^2\, \tau \int_{-\infty}^{\infty}\!\!\!\!d\eta_s e^{-p_T/T\cosh(y-\eta_s)} \mathscr{R}(T,\eta_s),
\end{equation}
with
\begin{align}
 \mathscr{R}(T,\eta_s) =& \ln \left[\frac{3.7388\, p_T\cosh(y-\eta_s)}{g^{2} T}\right]
 \frac{1}{p_T \cosh(y-\eta_s)}
 \nonumber \\
 &  \times \left[\frac{p_{T}^{2}}{2}-p_{T}^{2} \sinh ^{2}\left(y-\eta_{s}\right)\right] \frac{5\bar{\pi}}{2T}. 
\end{align}
The thermal particle spectra can be obtained by numerically integrating the above expressions over the space-time history of heavy-ion collisions along with the temperature profile of the expanding quark-gluon plasma.

\section{Results and discussions}
\label{results}

In this section, we calculate the spectra of thermal particles by employing the analytical solutions corresponding to the three approximations $\tau_\pi=$ const., $\tau_\pi \sim 1/T_{id}$ and $\tau_\pi \sim 1/T_{NS}$. Since third-order evolution is found to better reproduce the exact solution of kinetic theory compared to MIS and DNMR theories~\cite{Jaiswal:2013vta}, we consider coefficients corresponding to third-order theory 
(refer Table~\ref{coeff}) for allowed values of $\alpha$. Evolution of temperature of expanding hot QGP medium is determined from the expression of energy density, Eq.~(\ref{generic_energy}), and is given by
\begin{eqnarray} 
\label{TEMP}
 T(\bar{\tau}) &=& T_0 \left(\frac{\omega_0}{\omega}\right)^{\frac{1}{3}
 \left(|\Lambda|-\frac{k}{\gamma}\right)} e^{-\frac{1}{6\gamma}(\omega-\omega_0)} 
 \nonumber \\
&&\times \left(\frac{M_{k,m}(\omega)+\alpha W_{k,m}(\omega)}{M_{k,m}(\omega_0)
+\alpha W_{k,m}(\omega_0)}\right) ^{\frac{1}{3\gamma}}.
\end{eqnarray}
The expression for $T(\bar{\tau})$ for various approximations of $\tau_\pi$ is calculated by providing the parameters corresponding to each case from Table~\ref{T2}. The temperature profiles of attractor and repulsor solutions are obtained from Eq.~(\ref{TEMP}) in the limits $\alpha \rightarrow 0$ and $\alpha\rightarrow\infty$ respectively:
\begin{align}
 T_{\textrm{att}}(\bar{\tau}) &= T_0 \left(\frac{\omega_0}{\omega}\right)^
 {\frac{1}{3}\left(|\Lambda|-\frac{k}{\gamma}\right)} e^{-\frac{1}{6\gamma}(\omega-\omega_0)}\! 
 \left[\frac{M_{k,m}(\omega)}{M_{k,m}(\omega_0)}\right]^{\frac{1}{3\gamma}}, \label{T_att}\\
 T_{\textrm{repl}}(\bar{\tau})&= T_0 \left(\frac{\omega_0}{\omega}\right)^
 {\frac{1}{3}\left(|\Lambda|-\frac{k}{\gamma}\right)} e^{-\frac{1}{6\gamma}(\omega-\omega_0)}\! 
 \left[\frac{W_{k,m}(\omega)}{W_{k,m}(\omega_0)}\right]^{\frac{1}{3\gamma}}. \label{T_repl}
\end{align}
Next, we intend to study the thermal particle spectra in the presence of viscosity for the three approximations of $\tau_\pi$ by employing the temperature profiles calculated in this section. For this analysis, we take the initial conditions relevant to RHIC energies i.e., $T_0=360\,$MeV, $\tau_0=0.6\,$fm/c and the value of critical temperature $T_c$ is fixed as $155\,$MeV. The system is evolved till $\tau_f$ which is the time taken to attain $T_c$. Note that, since Eq.(\ref{TEMP}) depends on the parameter $\alpha$, we determine the value of $\tau_f$ for each case by varying $\alpha$. Table.~\ref{tau_f} shows the values of $\tau_f$ corresponding to analytical attractor and repulsor temperature profiles for the different approximations of $\tau_\pi$. $\tau_f$ corresponding to other allowed values of $\alpha$ are found to lie within these bounds.
%
\begin{table}[h!]
 \begin{center}
  \begin{tabular}{|c|c|c|c|c|}
   \hline 
   \phantom{$\Big|$} \phantom{$\Big|$} & $\,\,\tau_{\pi} = \textrm{const.}\,\,$  & $\,\,\tau_{\pi} \sim 1/T_{\textrm{id}}\,\,$  & $\,\,\tau_{\pi} \sim 1/T_{\textrm{NS}}\,\,$	 \\
   \hline
   \phantom{$\Big|$} Attractor \phantom{$\Big|$} & $\,10.15\,$ & $\, 8.46\,$  & $\,8.45\,$	\\
   \hline
   \phantom{$\Big|$} Repulsor  \phantom{$\Big|$} & $\,2.26\,$  & $\,1.49\,$   & $\,1.48\,$  \\
   \hline
   \end{tabular}
 \end{center}
\vspace*{-4mm}
\caption{$\tau_f$ values (in $fm/c$) corresponding to analytical attractor ($\alpha=0$) and repulsor ($\alpha=\infty$) temperature profiles in the three approximations of $\tau_{\pi}$.  }
\label{tau_f}
\vspace*{-2mm}
\end{table}

In order to calculate the ideal spectra (ideal rate, ideal evolution), we integrate the ideal contribution to thermal particle yield (Eq.~(\ref{dilid-yield}) \& Eq.~(\ref{phID-yield})) along with the ideal Bjorken solution $T_{\rm id}(\tau)=T_0\left(\frac{\tau_0}{\tau}\right)^{1/3}$. Here, the integration is carried out till $\tau_f=7.5\,$fm/c. For the ideal evolution of dileptons, we present an analytic expression which is obtained by integrating Eq.~(\ref{dilid-yield}) together with $T_{\rm id}(\tau)$
\begin{eqnarray}\label{dil-idEVO}
 \frac{d N_{l^{+}l^{-}}^{i}}{d M^2 d^2 p_{T} d y} &=& 6 \mathscr{F} 
 \frac{\tau_0^2}{z_0^6} \Bigg[z_T \Big[(4 z_T^3 + 32 z_T) K_0(z_T)\nonumber \\
 &&+ (z_T^4 + 16z_T^2 + 64) K_1(z_T) \Big]\Bigg]_{z_f}^{z_0},
\end{eqnarray}
where $z_0=m_T/T_0$ and $z_f=m_T/T_c$.\\

We analyze the effect of viscosity to thermal particle yields by constructing the following ratios:
\begin{eqnarray}
 R_{l^{+}l^{-}} &=& \left(\frac{d N_{l^{+}l^{-}}}{d M^2 d^2 p_{T} d y}\right) \!\Big/\!
 \left(\frac{d N_{l^{+}l^{-}}^i}{d M^2 d^2 p_{T} d y}\right) \,,
 \label{ratio_dilepton} \\
 R_{\gamma} &=& \left(\frac{d N_{\gamma}}{d^{2} p_{T} d y}\right) \!\Big/\!
 \left(\frac{d N_{\gamma}^{i}}{d^{2} p_{T} d y}\right). 
 \label{ratio_photon}
\end{eqnarray}
Here, the ratios $R_{l^{+}l^{-}}$ and $R_{\gamma}$ represent the relative magnitude of viscous corrections to ideal thermal dilepton and photon evolutions respectively. We plot these ratios as a function of transverse momentum of the particles by varying the parameter $\alpha$. We present our results for the midrapidity region of the particles i.e., for $y=0$.
 
\begin{figure}[t!]
 \includegraphics[width=\linewidth]{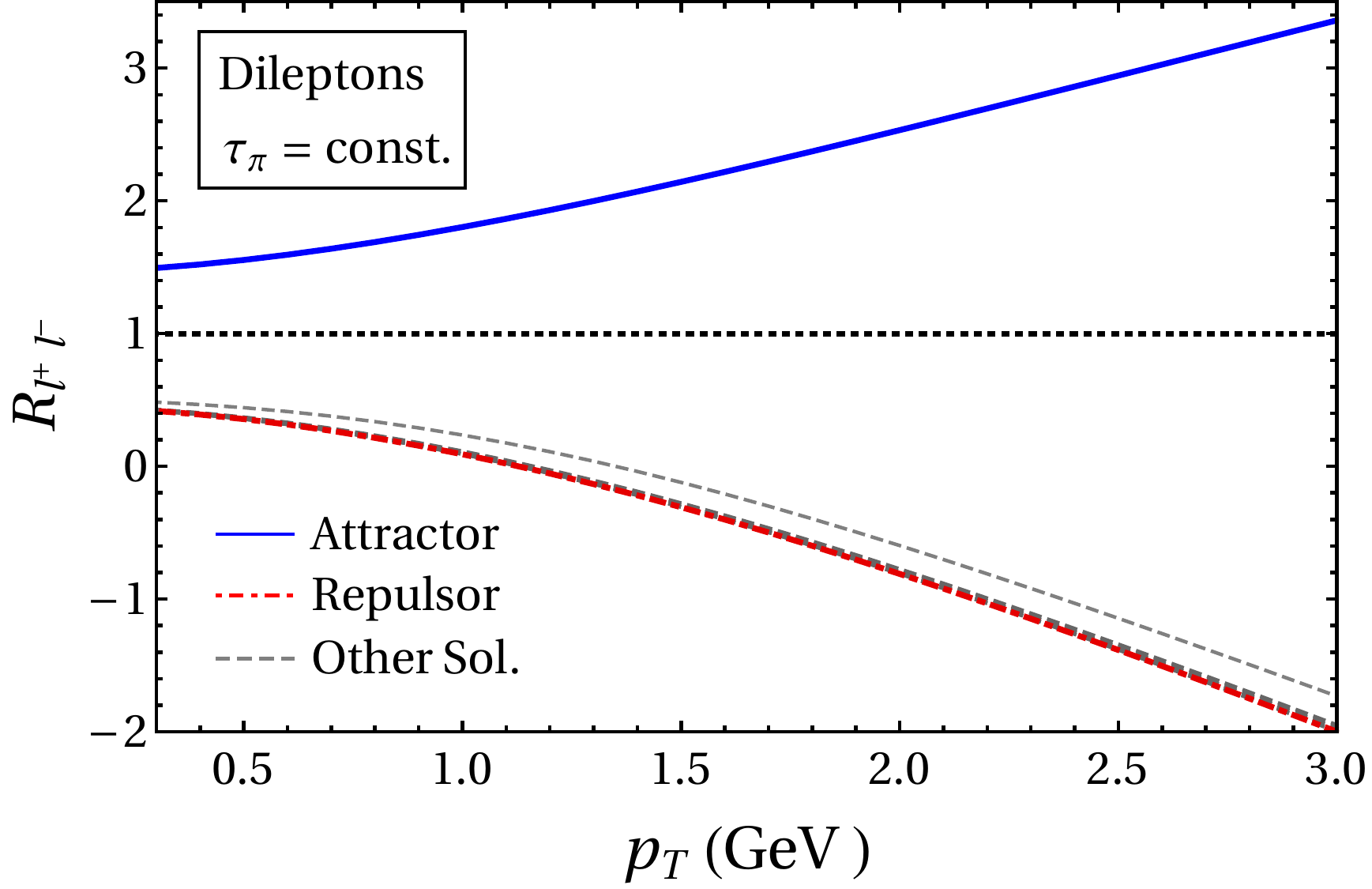}
 \vspace{-0.6cm}
 \caption{Ratio of viscous to ideal dilepton spectra, Eq.~\eqref{ratio_dilepton}, for $\tau_\pi = {\textrm{const}}$. Here, solid blue curve and dot-dashed red curve indicates the spectra corresponding to attractor and repulsor solution, respectively. The black dotted line corresponds to the ratios equal to 1, i.e., the ideal case. The dashed grey curves represents various initial conditions in the viscous evolution governed by values of $\alpha$ ranging from $10$ to $5000$. Here we consider the dilepton invariant mass, $M=1$~GeV.}
\label{Rdil-cnstRT}
 \vspace{-0.2cm}
\end{figure}

\begin{figure}[t!]
 \includegraphics[width=\linewidth]{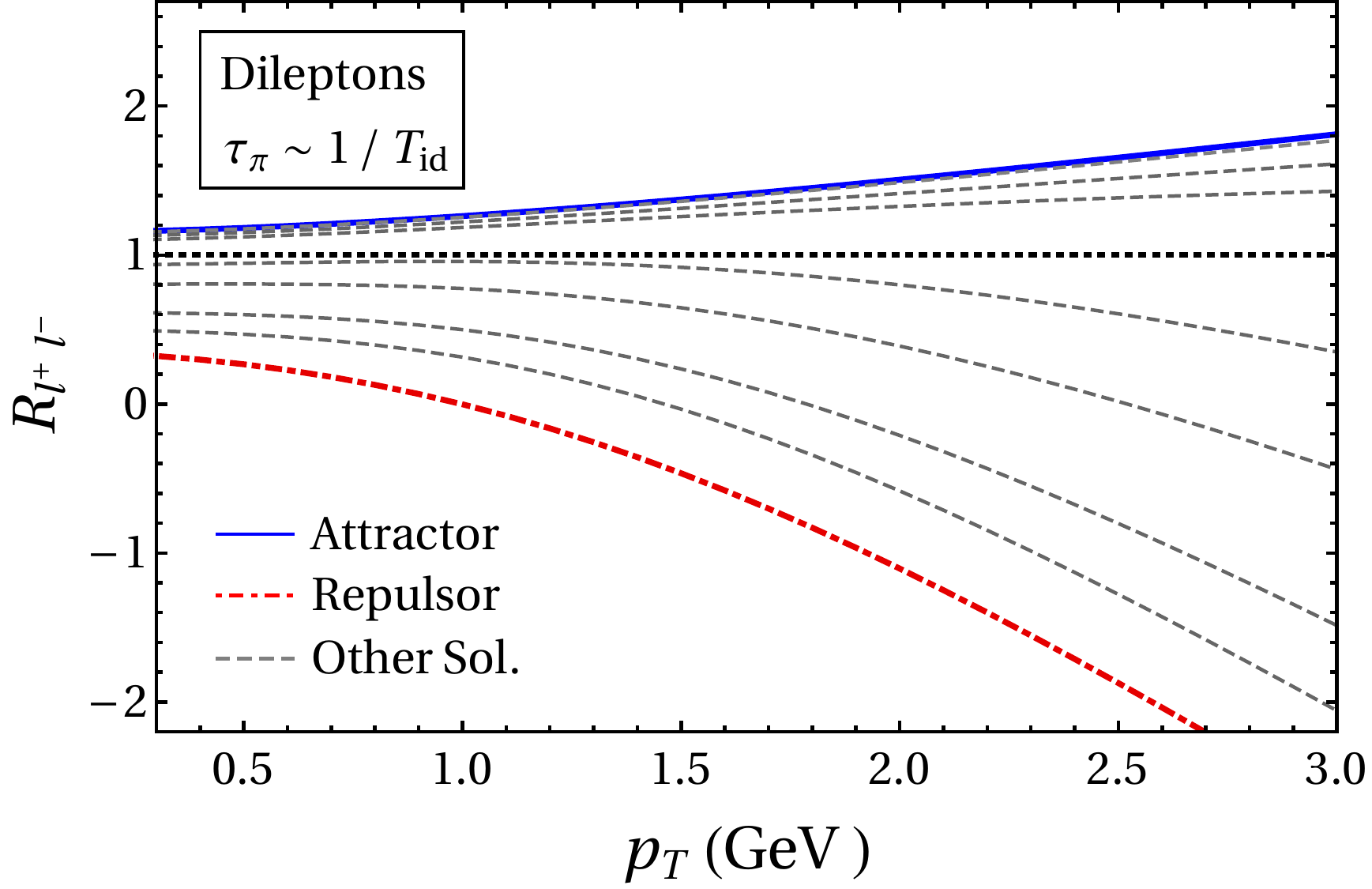}
 \vspace{-0.6cm}
 \caption{Same as Fig.~\ref{Rdil-cnstRT} but for $\tau_\pi \sim 1/T_{\textrm{id}}$.}
\label{Rdil-id}
 \vspace{-0.2cm}
\end{figure}

\begin{figure}[t!]
 \includegraphics[width=\linewidth]{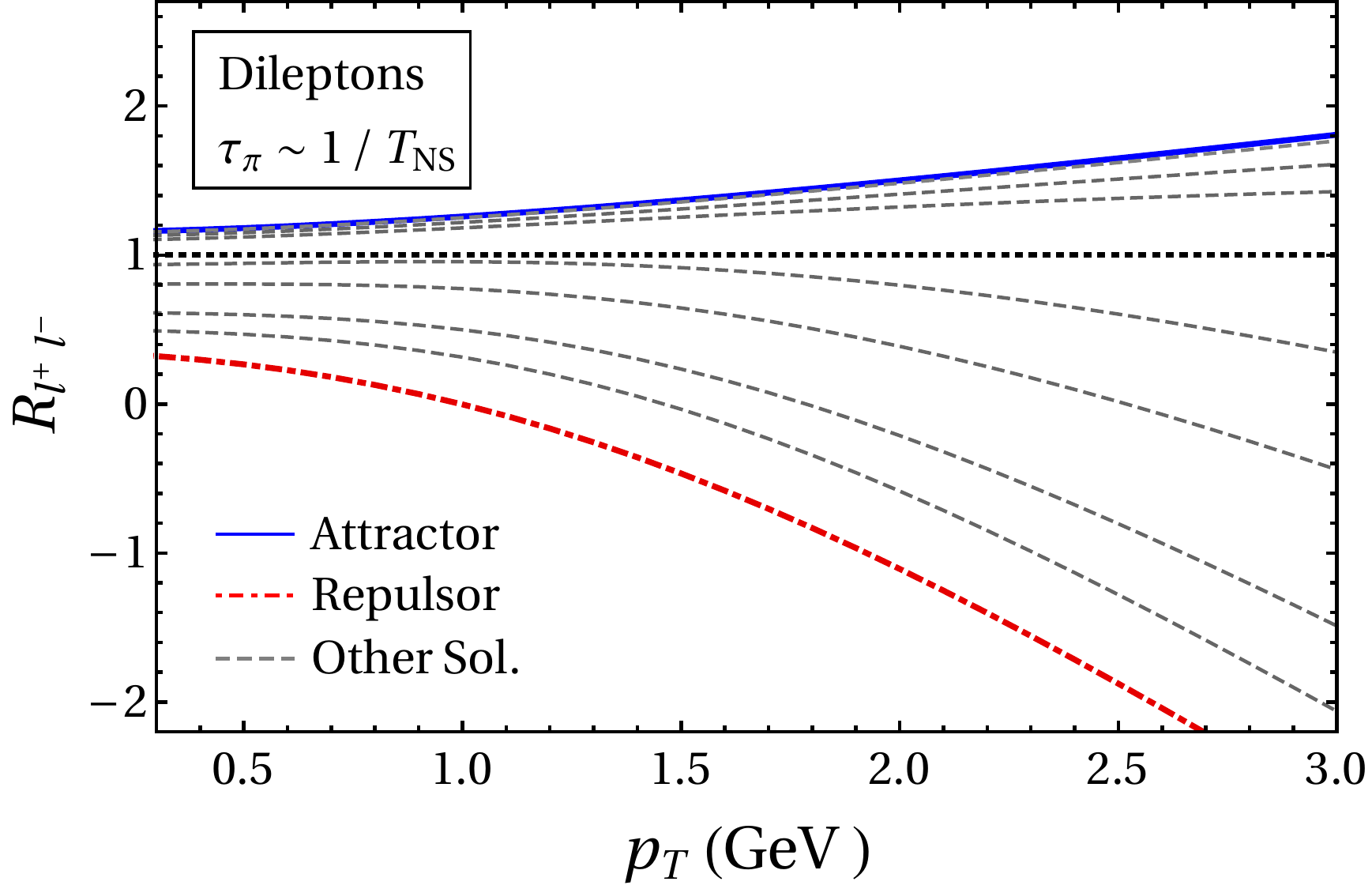}
 \vspace{-0.6cm}
 \caption{Same as Fig.~\ref{Rdil-cnstRT} but for $\tau_\pi \sim 1/T_{\textrm{NS}}$.}
\label{Rdil-NS}
 \vspace{-0.2cm}
\end{figure}

\begin{figure}[t!]
 \includegraphics[width=\linewidth]{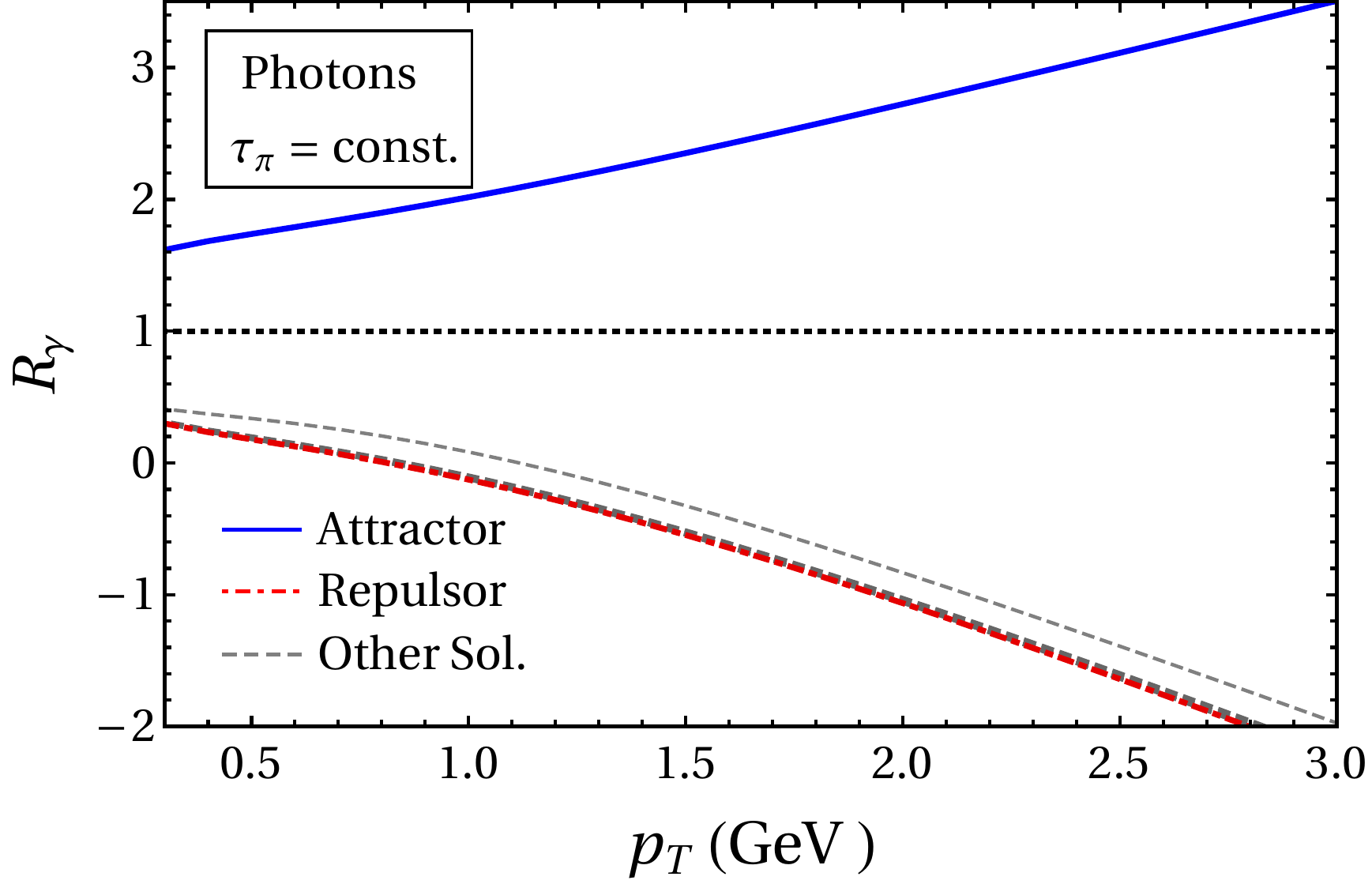}
 \vspace{-0.6cm}
 \caption{Ratio of viscous to ideal photon spectra, Eq.~\eqref{ratio_photon}, for $\tau_\pi = {\textrm{const}}$. Here, solid blue curve and dot-dashed red curve indicates the spectra corresponding to attractor and repulsor solution, respectively. The black dotted line corresponds to the ratios equal to 1, i.e., the ideal case. The dashed grey curves represents various initial conditions in the viscous evolution governed by values of $\alpha$ ranging from $10$ to $5000$.}
\label{Rph-cnstRT}
 \vspace{-0.2cm}
\end{figure}

\begin{figure}[t!]
 \includegraphics[width=\linewidth]{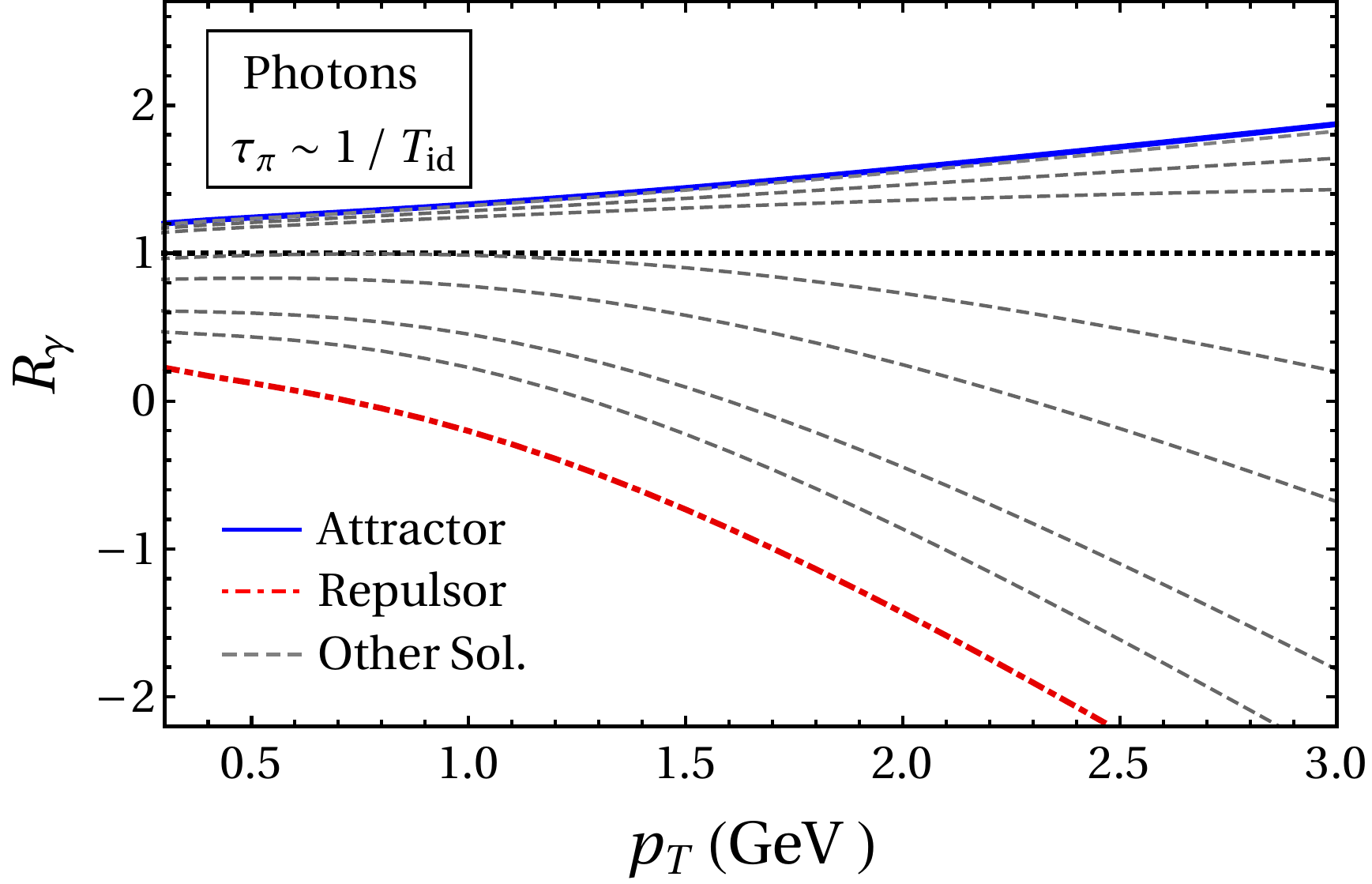}
 \vspace{-0.6cm}
 \caption{Same as Fig.~\ref{Rph-cnstRT} but for $\tau_\pi \sim 1/T_{\textrm{id}}$.}
\label{Rph-id}
 \vspace{-0.2cm}
\end{figure}

\begin{figure}[t!]
 \includegraphics[width=\linewidth]{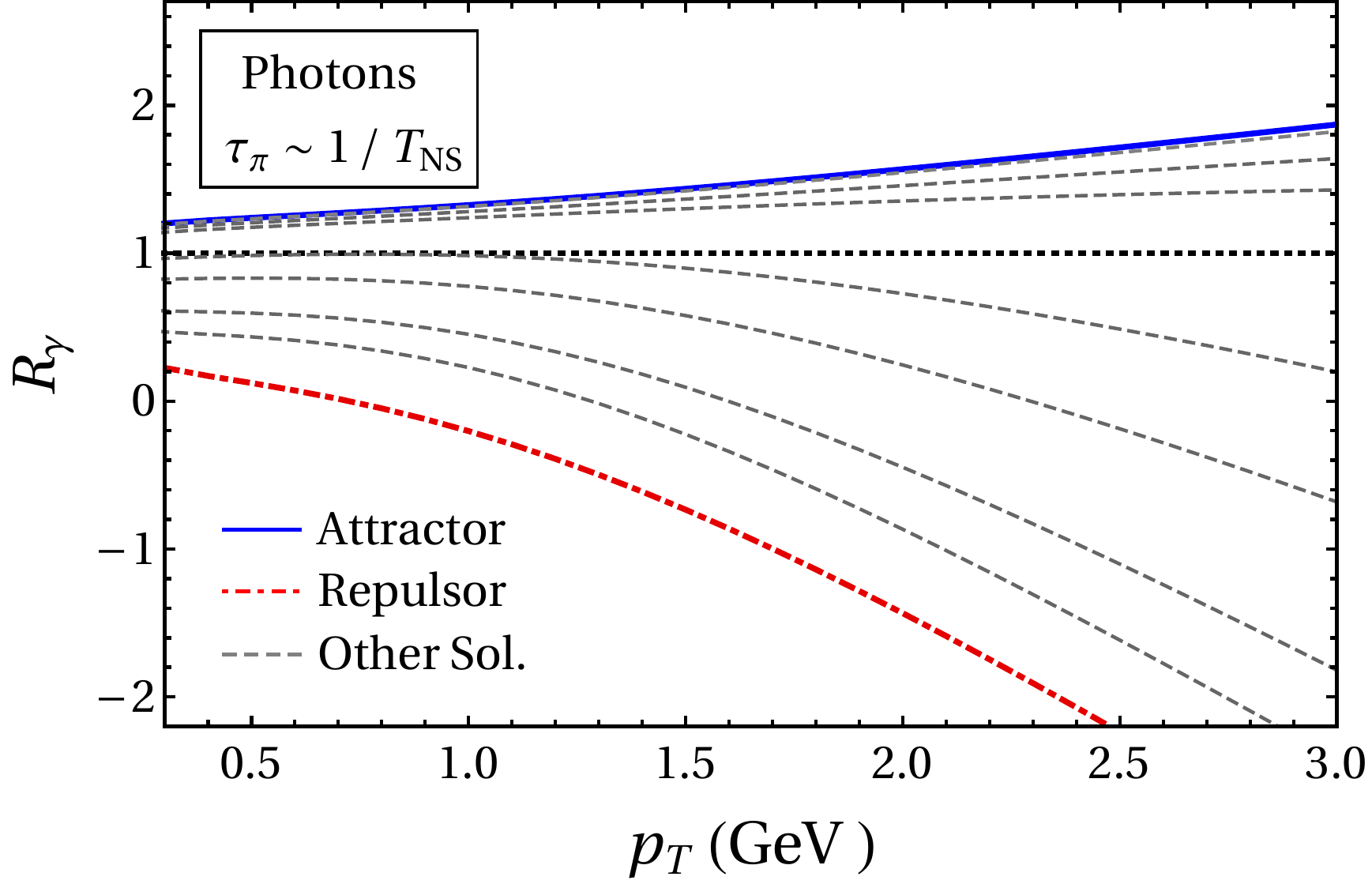}
 \vspace{-0.6cm}
 \caption{Same as Fig.~\ref{Rph-cnstRT} but for $\tau_\pi \sim 1/T_{\textrm{NS}}$.}
\label{Rph-NS}
 \vspace{-0.2cm}
\end{figure}

In Figs.~\ref{Rdil-cnstRT}~to~\ref{Rph-NS}, solid blue curve and dot-dashed red curve indicates the spectra corresponding to attractor and repulsor solution, respectively. The black dotted line corresponds to the ratios being unity, i.e., the ideal case. The dashed grey curves represent various initial conditions in the viscous evolution governed by values of $\alpha$ ranging from $10$ to $5000$. It is observed that the particle yields are maximum for the attractor and minimum for the repulsor. In Fig.~\ref{Rdil-cnstRT}, we plot the ratio of viscous corrections to \textit{ideal} thermal dilepton spectra for constant $\tau_\pi$. The viscous contribution for non-zero $\alpha$ values tends to approach the repulsor one, indicating a large suppression of yields even for a small increment in $\alpha$. 

We present the equivalent plots for the other two approximate solutions, $\tau_\pi \sim 1/T_{\textrm{id}}$ and $\sim 1/T_{\textrm{NS}}$ in Figs.~\ref{Rdil-id} and \ref{Rdil-NS} respectively. It is crucial to note that the viscous contributions to the spectra appears to be almost identical for both these approximations. There is an overall increment in the viscous contributions throughout the $p_T$ regime for the ideal and Navier-Stokes relaxation time approximations with small values of $\alpha$ (for initializations near the attractor). Also the curves approaches the repulsor with increase in $\alpha$ which indicates the suppression in the yield for large $\alpha$. We note that the particle yield remains maximum for the attractor, as was seen in Fig.~\ref{Rdil-cnstRT}.

Further, we study the total photon spectra normalized by the \textit{ideal} case for the three approximations in $\tau_\pi$. Figs.~\ref{Rph-cnstRT},~\ref{Rph-id} and~\ref{Rph-NS} display the ratio $R_{\gamma}$ as a function of transverse momentum of photons for $\tau_\pi = {\textrm{const}}.$, $\tau_\pi \sim 1/T_{\textrm{id}}$ and $\tau_\pi \sim 1/T_{\textrm{NS}}$ respectively. Here, it should be noted that the ratio of viscous corrections to the ideal photon spectra is almost identical to those as observed in $R_{l^{+}l^{-}}$ plots. Large $\alpha$ values suppress the thermal photon spectra over the entire $p_T$ regime.    

It is important to note that for large values of $\alpha$, the viscous spectra becomes negative even at small $p_T$, as seen Figs.~(\ref{Rdil-cnstRT}~-\ref{Rph-NS}). This is due to the fact that viscous corrections to the distribution function become large and negative as the initial condition approaches that of repulsor; see Fig.~\ref{stream_plot}.

\begin{figure}[t!]
\includegraphics[width=\linewidth]{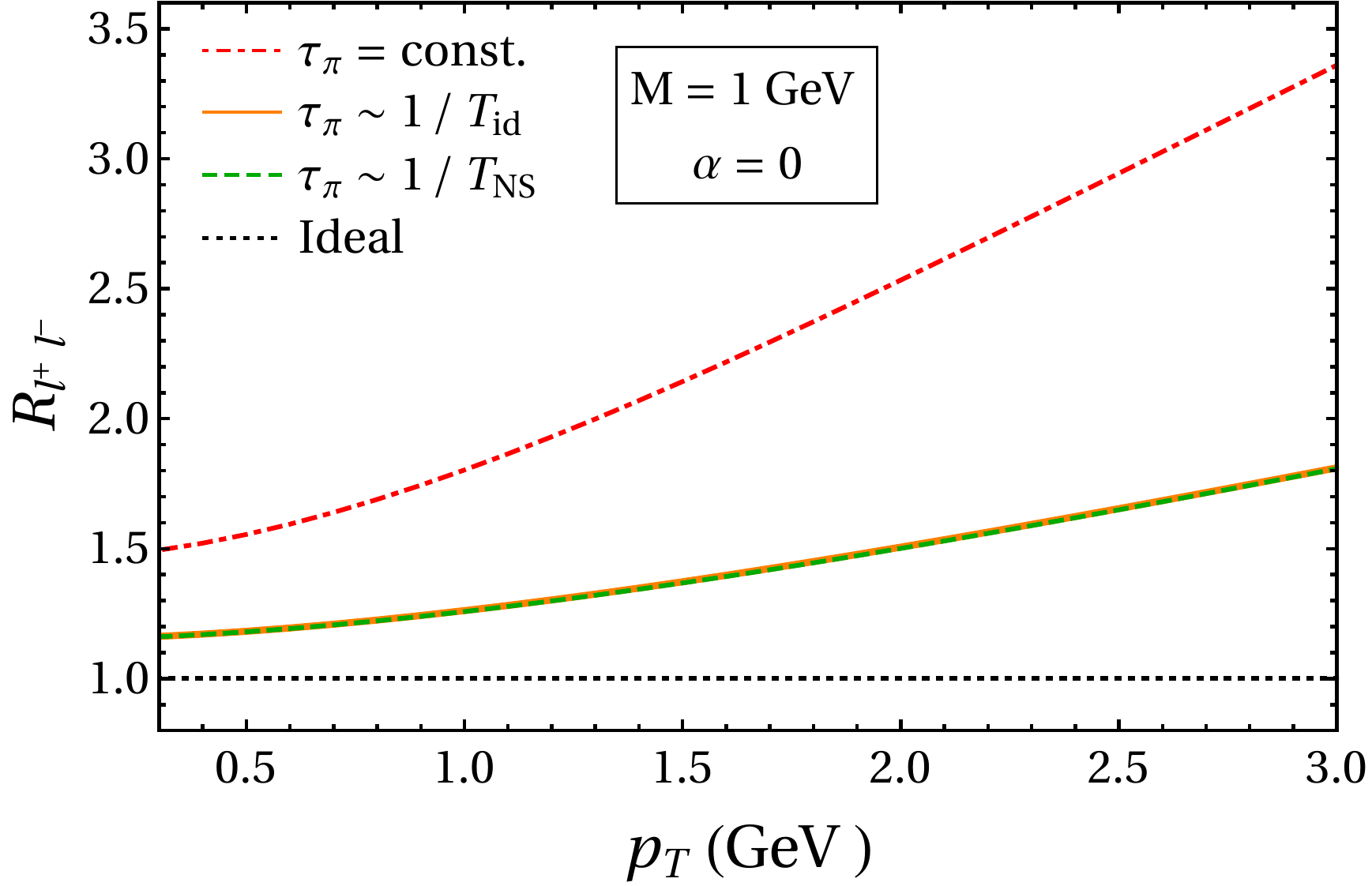}
\vspace{-0.6cm}
 \caption{Ratio of viscous to ideal dilepton spectra, Eq.~\eqref{ratio_dilepton}, corresponding to attractor solutions, Eq.~\eqref{T_att}, for $\tau_\pi = {\textrm{const}}$, $\tau_\pi \sim 1/T_{\textrm{id}}$ and $\tau_\pi \sim 1/T_{\textrm{NS}}$. Here we consider the 
 dilepton invariant mass, $M=1$~GeV.}
 \label{Rdil-alpha}
\end{figure}

\begin{figure}[t!]
\includegraphics[width=\linewidth]{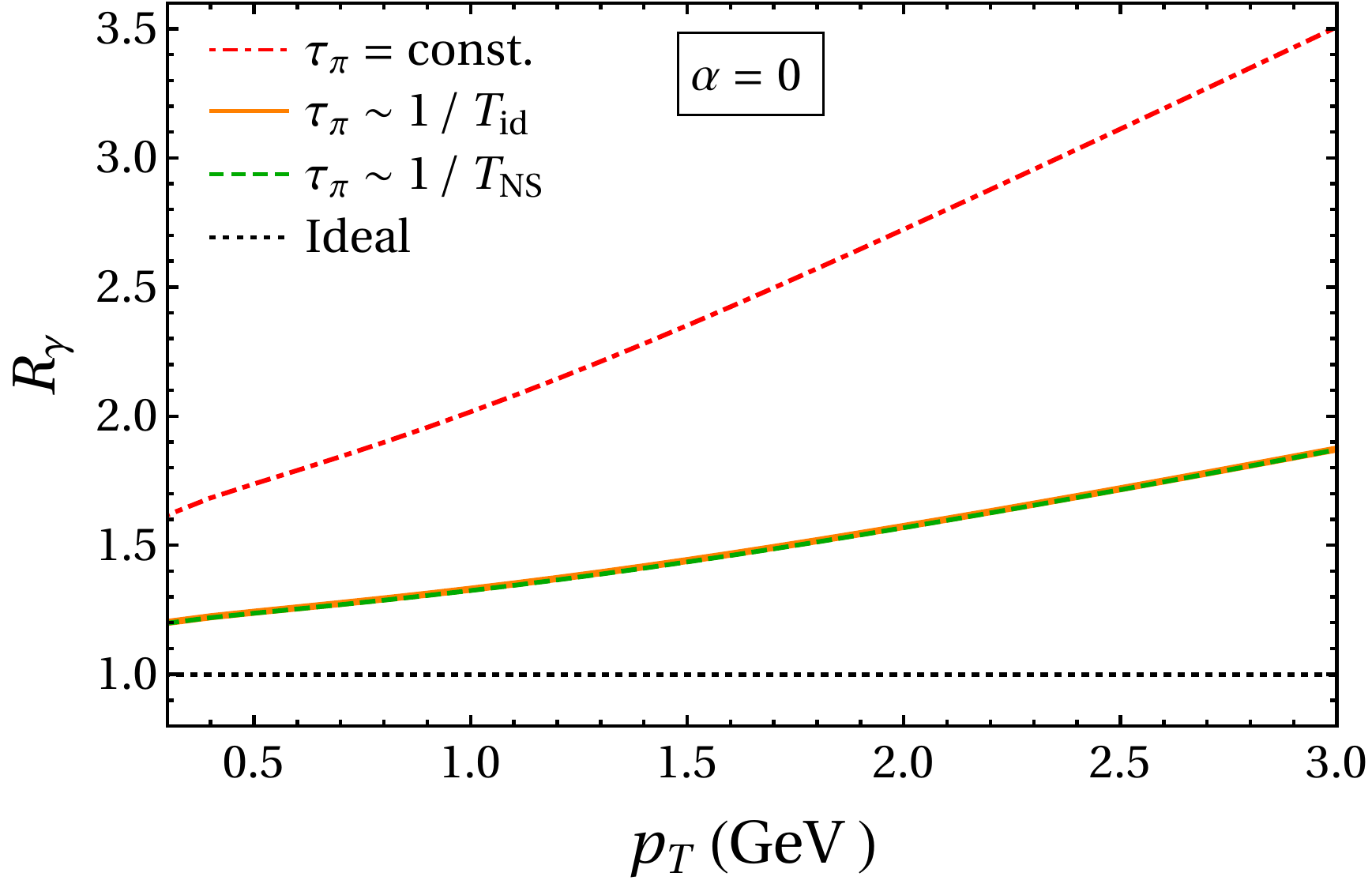}
\vspace{-0.6cm}
 \caption{Ratio of viscous to ideal photon spectra, Eq.~\eqref{ratio_photon}, corresponding to attractor solutions, Eq.~\eqref{T_att}, for $\tau_\pi = {\textrm{const}}$, $\tau_\pi \sim 1/T_{\textrm{id}}$ and $\tau_\pi \sim 1/T_{\textrm{NS}}$.}
 \label{Rph-alpha}
\end{figure}
 
Figs.~\ref{Rdil-alpha} and \ref{Rph-alpha} show comparison between the ratios of viscous to ideal dilepton and photon spectra corresponding to attractor solutions, given by Eq.~\eqref{T_att}, obtained from various approximations: for $\tau_\pi = {\textrm{const}}$, $\tau_\pi \sim 1/T_{\textrm{id}}$ and $\tau_\pi \sim 1/T_{\textrm{NS}}$. As mentioned earlier, the corrections in the case of $\tau_\pi \sim 1/T_{\textrm{id}}$ are found to be comparable with that of $\tau_\pi \sim 1/T_{\textrm{NS}}$. Also, these results display that the spectra corresponding to $\tau_\pi = {\textrm{const}}$  differ significantly from the other two cases. This difference is nominal in the low $p_T$ regime and tend to increase with $p_T$.

\section{Summary and outlook}
\label{summary}

In this article, we considered thermal particle production within the framework of relativistic viscous hydrodynamics and employed recently obtained analytical solutions of higher-order viscous hydrodynamics. Demanding positivity and reality of energy density throughout the evolution, we constrained the allowed region in the basin of attraction. We then calculated the non-equilibrium correction to both dilepton and photon production rates by employing viscous correction to the distribution function obtained using Chapman-Enskog like expansion of the Boltzmann equation in the relaxation-time approximation. Further, we studied the effect of hydrodynamic evolution corresponding to attractor and repulsor solutions on the thermal particle spectra. We have found that the viscous corrections enhance the thermal particle spectra for small value value of parameter $\alpha$ (which controls initial conditions) and suppress the spectra for large $\alpha$. Moreover, the yields corresponding to attractor gets the maximum enhancement and the ones corresponding to repulsor suffers maximum suppression. For $\tau_\pi^{\textrm{id}}$ and $\tau_\pi^{\textrm{NS}}$, it can be observed that the viscous contributions to spectra are nearly identical while for $\tau_\pi^{\textrm{const}}$, the corrections are less. 

Looking forward, it will be interesting to consider pre-equilibrium dynamics in the current framework and study this effect on the dilepton and photon spectra. Indeed, it has been recently shown that intermediate mass dileptons can act as pre-equilibrium probes in heavy ion collisions~\cite{Coquet:2021cuv}. We leave this for future study.


\begin{acknowledgments}
A.J. is supported in part by the DST-INSPIRE faculty award under Grant No. DST/INSPIRE/04/2017/000038.
\end{acknowledgments}

\bibliography{references}
\end{document}